\documentclass[twocolumn]{aastex63}

\usepackage{graphicx}

\begin{document}

\title{Temperature structures of embedded disks: young disks in Taurus are warm}

\correspondingauthor{Merel L.R. van 't Hoff}
\email{mervth@umich.edu}

\author{Merel L.R. van 't Hoff}
\affil{Leiden Observatory, Leiden University, P.O. box 9513, 2300 RA Leiden, The Netherlands}
\affil{Department of Astronomy, University of Michigan, 500 Church Street, Ann Arbor, MI 48109, USA}

\author{Daniel Harsono}
\affil{Leiden Observatory, Leiden University, P.O. box 9513, 2300 RA Leiden, The Netherlands}
\affil{Institute of Astronomy and Astrophysics, Academia Sinica, No. 1, Sec. 4, Roosevelt Road, Taipei 10617, Taiwan, R. O. C.}

\author{John J. Tobin}
\affil{National Radio Astronomy Observatory, 520 Edgemont Rd., Charlottesville, VA 22903, USA}

\author{Arthur D. Bosman}
\affil{Leiden Observatory, Leiden University, P.O. box 9513, 2300 RA Leiden, The Netherlands}
\affil{Department of Astronomy, University of Michigan, 500 Church Street, Ann Arbor, MI 48109, USA}

\author{Ewine F. van Dishoeck}
\affil{Leiden Observatory, Leiden University, P.O. box 9513, 2300 RA Leiden, The Netherlands}
\affil{Max-Planck-Institut f\"ur Extraterrestrische Physik, Giessenbachstrasse 1, D-85748 Garching bei M\"unchen, Germany}

\author{Jes K. J{\o}rgensen}
\affil{Niels Bohr Institute, University of Copenhagen, {\O}ster Voldgade 5-7, DK-1350 Copenhagen K., Denmark}

\author{Anna Miotello} 
\affil{European Southern Observatory, Karl-Schwarzschild-Str. 2, D-85748 Garching bei M\"unchen, Germany}

\author{Nadia M. Murillo} 
\affil{Leiden Observatory, Leiden University, P.O. box 9513, 2300 RA Leiden, The Netherlands}
\affil{The Institute of Physical and Chemical Research (RIKEN), 2-1, Hirosawa, Wako-shi, Saitama 351-0198, Japan}

\author{Catherine Walsh}
\affil{School of Physics and Astronomy, University of Leeds, Leeds LS2 9JT, UK}


\begin{abstract} 

\noindent The chemical composition of gas and ice in disks around young stars set the bulk composition of planets. In contrast to protoplanetary disks (Class II), young disks that are still embedded in their natal envelope (Class 0 and I) are predicted to be too warm for CO to freeze out, as has been confirmed observationally for L1527 IRS. To establish whether young disks are generally warmer than their more evolved counterparts, we observed five young (Class 0/I and Class I) disks in Taurus with the Atacama Large Millimeter/submillimeter Array (ALMA), targeting C$^{17}$O $2-1$, H$_2$CO $3_{1,2}-2_{1,1}$, HDO $3_{1,2}-2_{2,1}$ and CH$_3$OH $5_K-4_K$ transitions at $0.48^{\prime\prime} \times 0.31^{\prime\prime}$ resolution. The different freeze-out temperatures of these species allow us to derive a global temperature structure. C$^{17}$O and H$_2$CO are detected in all disks, with no signs of CO freeze-out in the inner $\sim$100 au, and a CO abundance close to $\sim$10$^{-4}$. H$_2$CO emission originates in the surface layers of the two edge-on disks, as witnessed by the especially beautiful V-shaped emission pattern in IRAS~04302+2247. HDO and CH$_3$OH are not detected, with column density upper limits more than 100 times lower than for hot cores. Young disks are thus found to be warmer than more evolved protoplanetary disks around solar analogues, with no CO freeze-out (or only in the outermost part of $\gtrsim$100 au disks) or CO processing. However, they are not as warm as hot cores or disks around outbursting sources, and therefore do not have a large gas-phase reservoir of complex molecules.

\vspace{1cm}

\end{abstract}



\section{Introduction} \label{sec:intro}

Disks around young stars provide the material from which planets form. Knowledge of their physical and chemical structure is therefore crucial for understanding planet formation and composition. The physics of protoplanetary disks has been studied in great detail, both using observations of individual objects \citep[e.g.,][]{vanZadelhoff2001,Andrews2010,Andrews2018,Schwarz2016} as well as through surveys of star-forming regions \citep[e.g.,][]{Ansdell2016,Ansdell2017,Barenfeld2016,Pascucci2016,Cox2017,Ruiz-Rodriguez2018,Cieza2019}. Molecular line observations require more telescope time than continuum observations, hence studies of the chemical structure generally target individual disks or small samples of bright disks \citep[e.g.,][]{Dutrey1997,Thi2004,Oberg2010,Cleeves2015,Huang2017}. The picture that is emerging for the global composition of Class II disks around solar analogues is that they have a large cold outer region ($T \lesssim$ 20 K) where CO is frozen out in the disk midplanes \citep[e.g.,][]{Aikawa2002,Qi2013,Qi2015,Qi2019,Mathews2013,Dutrey2017}. 

However, it is now becoming clear that planet formation already starts when the disk is still embedded in its natal envelope. Grain growth has been observed in Class 0 and I sources and even larger bodies may have formed before the envelope has fully dissipated \citep[e.g.,][]{Kwon2009,Jorgensen2009,Miotello2014,ALMAPartnership2015,Harsono2018}. Furthermore, the dust mass of Class II disks seems insufficient to form the observed exoplanet population, but Class 0 and I disks are massive enough \citep{Manara2018,Tychoniec2020}. Young embedded disks thus provide the initial conditions for planet formation, but unlike their more evolved counterparts, their structure remains poorly characterized. 

A critical property is the disk temperature structure because this governs disk evolution and composition. For example, temperature determines whether the gas is susceptible to gravitational instabilities (see, e.g., a review by \citealt{Kratter2016}), a potential mechanism to form giant planets, stellar companions and accretion bursts \citep[e.g.,][]{Boss1997,Boley2009,Vorobyov2009,Tobin2016a}. In addition, grain growth is thought to be enhanced in the region where water freezes out from the gas phase onto the dust grains, the water snowline ($T \sim$100--150 K; e.g., \citealt{Stevenson1988,Schoonenberg2017,Drazkowska2017}).

Moreover, freeze out of molecules as the temperature drops below their species-specific freeze-out temperature sets the global chemical composition of the disk. This sequential freeze-out causes radial gradients in molecular abundances and elemental ratios (like the C/O ratio, e.g., \citealt{Oberg2011}). In turn, the composition of a planet then depends on its formation location in the disk \citep[e.g.,][]{Madhusudhan2014,Walsh2015,Ali-Dib2017,Cridland2019}. Finally, the formation of high abundances of complex molecules starts from CO ice \citep[e.g.,][]{Tielens1982,Garrod2006,Cuppen2009,Chuang2016} and COM formation will thus be impeded during the disk stage if the temperature is above the CO freeze-out temperature ($T \gtrsim$~20~K). Whether young disks are warm ($T \gtrsim$~20~K, i.e., warmer than the CO freeze-out temperature) or cold (i.e., have a large region where $T \lesssim$~20~K and CO is frozen out) is thus a simple, but crucial question.  

\begin{deluxetable*}{l l c c c c c c c c c c c}
\tablecaption{Overview of source properties.
\label{tab:SourceOverview}}
\tablewidth{100pt}
\tabletypesize{\scriptsize}
\tablehead{
\colhead{Source name} \vspace{-0.3cm} & \colhead{Other name} & \colhead{R.A.\tablenotemark{a}} & \colhead{Decl.\tablenotemark{a}} & \colhead{Class} & \colhead{$\mathbf{T_{\rm{bol}}}$} & \colhead{$L_{\rm{bol}}$} & \colhead{$M_{\ast}$} & \colhead{$M_{\rm{env}}$} & \colhead{$M_{\rm{disk}}$} & \colhead{$R_{\rm{disk}}$} & \colhead{$i$} & \colhead{Refs\tablenotemark{b}} \\ 
\colhead{(IRAS)} \vspace{-0.5cm} & \colhead{} & \colhead{(J2000)} & \colhead{(J2000)} & \colhead{} & \colhead{(K)} & \colhead{($L_{\odot}$)} & \colhead{($M_{\odot}$)} & \colhead{($M_{\odot}$)}  & \colhead{($M_{\odot}$)} & \colhead{(au)} & \colhead{(deg)} & \colhead{} \\ 
} 
\startdata 
04016+2610 & L1489 IRS     & 04:04:43.1 & +26:18:56.2 & I      & 226 & 3.5              & 1.6           & 0.023    & 0.0071  & 600 & 66 & 1--4 \\
04302+2247 & Butterfly star & 04:33:16.5 & +22:53:20.4 & I/II & 202 &0.34--0.92 & 0.5\tablenotemark{c}  & 0.017 & 0.11 & 244 & $>$76  & 3,5,9 \\
04365+2535 & TMC1A          & 04:39:35.2 & +25:41:44.2 & I     & 164 &2.5             & 0.53--0.68 & 0.12      & 0.003--0.03     & 100            & 50 & 1,6--8 \\
04368+2557 & L1527 IRS     & 04:39:53.9 & +26:03:09.5 & 0/I  & 59 &1.9--2.75      & 0.19--0.45   & 0.9--1.7      & 0.0075  & 75--125    & 85   & 9--14 \\
04381+2540 & TMC1            & 04:41:12.7 & +25:46:34.8 & I       & 171 & 0.66--0.9     & 0.54        & 0.14    & 0.0039  & 100            & 55     & 1,6,10 \\
\enddata
\vspace{-0.5cm}
\begin{flushleft}
\tablecomments{All values presented in this table are from the literature listed in footnote b.}
\vspace{-0.2cm}
\tablecomments{TMC1 is resolved here for the first time as a binary. The literature values in this table are derived assuming a single source. }
\vspace{-0.2cm}
\tablenotetext{a}{Peak of the continuum emission, except for TMC1 where the phase center of the observations is listed. The coordinates of the two sources TMC1-E and TMC1-W are R.A. = 04:41:12.73, Decl = +25:46:34.76 and R.A. = 04:41:12.69, Decl = +25:46:34.73, respectively.}
\vspace{-0.15cm}
\tablenotetext{b}{References. (1) \citet{Green2013}, (2) \citet{Yen2014}, (3) \citet{Sheehan2017}, (4) \citet{Sai2020}, (5) \citet{Wolf2003}, (6) \citet{Harsono2014}, (7) \citet{Aso2015}, (8) Harsono et al., submitted (9) \citet{Motte2001}, (10) \citet{Kristensen2012}, (11) \citet{Tobin2008}, (12) \citet{Tobin2013}, (13) \citet{Oya2015} (14) \citet{Aso2017}.} 
\vspace{-0.2cm}
\tablenotetext{c}{Not a dynamical mass.}
\end{flushleft}
\end{deluxetable*}

Keplerian disks are now detected around several Class 0 and I sources \citep[e.g.,][]{Brinch2007,Tobin2012,Murillo2013,Yen2017}, but most research has focused on disk formation, size and kinematics \citep[e.g.,][]{Yen2013,Ohashi2014,Harsono2014}, or the chemical structure at the disk-envelope interface \citep[e.g.,][]{Sakai2014a,Murillo2015,Oya2016}. Only a few studies have examined the disk physical structure, and only for one particular disk, L1527~IRS. \citet{Tobin2013} and \citet{Aso2017} modeled the radial density profile and \citet{vantHoff2018b} studied its temperature profile based on optically thick $^{13}$CO and C$^{18}$O observations. The latter study showed the importance of disentangling disk and envelope emission and concluded that the entire L1527 disk is likely too warm for CO freeze-out, in agreement with model predictions \citep[e.g.,][]{Harsono2015}, but in contrast to observations of T Tauri disks. 

Another important question in regard of the composition of planet-forming material is the CO abundance. The majority of protoplanetary disks have surprisingly weak CO emission, even when freeze-out and isotope-selective photodissociation are taken into account \citep[e.g.,][]{Ansdell2016,Miotello2017,Long2017}. Based on gas masses derived from HD line fluxes \citep{Favre2013,McClure2016,Schwarz2016,Kama2016} and mass accretion rates \citep{Manara2016} the low CO emission seems to be the result of significant CO depletion (up to two orders of magnitude below the ISM abundance of $\sim$10$^{-4}$ with respect to H$_2$). 

Several mechanisms have been discussed in the literature, either focusing on the chemical conversion of CO into less volatile species \citep[e.g.,][]{Bergin2014,Eistrup2016,Schwarz2018,Schwarz2019,Bosman2018}, or using dust growth to sequester CO ice in the disk midplane \citep[e.g.,][]{Xu2017,Krijt2018}. Observations of CO abundances in younger disks can constrain the timescale of the CO depletion process. Observations of $^{13}$CO and C$^{18}$O toward the embedded sources TMC1A and L1527 are consistent with an ISM abundance \citep{Harsono2018,vantHoff2018b}. Recent work by \citet{Zhang2020} also found CO abundances consistent with the ISM abundance for three young disks in Taurus with ages upto $\sim$ 1 Myr using optically thin $^{13}$C$^{18}$O emission. Since the 2-3 Myr old disks in Lupus and Cha I show CO depletion by a factor 10--100 \citep{Ansdell2016}, these results suggest that the CO abundance decreases by a factor of ten within 1 Myr. On the other hand, \citet{Bergner2020} found C$^{18}$O abundances a factor of ten below the ISM value in two Class I sources in Serpens. 

In this paper we present ALMA observations of C$^{17}$O toward five young disks in Taurus to address the questions whether young disks are generally too warm for CO freeze-out and whether there is significant CO processing. The temperature profile is further constrained by H$_2$CO observations as this molecule freezes out around $\sim$70 K. Although chemical models often assume a binding energy of 2050 K \citep[e.g.,][]{Garrod2006,McElroy2013}, laboratory experiments have found binding energies ranging between 3300--3700 K depending on the ice surface \citep{Noble2012}. These latter values suggest H$_2$CO freeze-out temperatures between $\sim$70--90 K for disk-midplane densities ($\sim$10$^{8}-10^{10}$ cm$^{-3}$) instead of $\sim$50~K. Experiments by \citet{Fedoseev2015} are consistent with the lower end of binding energies found by \citet{Noble2012}, so we adopt a freeze-out temperature of 70 K for H$_2$CO. An initial analysis of these observations were presented in \citet[PhD thesis]{vantHoff2019}.

In addition, HDO and CH$_3$OH observations are used to probe the $\gtrsim100-150$~K region and to determine whether complex molecules can be observed in these young disks, as shown for the disk around the outbursting young star V883 Ori \citep{vantHoff2018c,Lee2019}. In contrast, observing complex molecules has turned out to be very difficult in mature protoplanetary disks. So far, only CH$_3$CN has been detected in a sample of disks, and CH$_3$OH and HCOOH have been detected in TW Hya \citep{Oberg2015,Walsh2016,Favre2018,Bergner2018,Loomis2018,Carney2019}. 

The observations are described in Sect.~\ref{sec:Observations}, and the resulting C$^{17}$O and H$_2$CO images are presented in Sect.~\ref{sec:Results}. This section also describes the non-detections of HDO and CH$_3$OH. The temperature structure of the disks is examined in Sect.~\ref{sec:Analysis} based on the C$^{17}$O and H$_2$CO observations and radiative transfer modeling. The result that the young disks in this sample are warm with no significant CO freeze out or CO processing is discussed in Sect.~\ref{sec:Discussion} and the conclusions are summarized in Sect.~\ref{sec:Conclusions}.


\begin{table*}
\caption{Observed fluxes for the 1.3 mm continuum and molecular lines.}
\label{tab:Flux} 
\centering
\begin{footnotesize}
\begin{tabular}{l c c c c c c}
    \hline\hline
    \\[-.3cm]
    Source & $F_{\mathrm{peak}}$ (1.3 mm) & $F_{\mathrm{int}}$ (1.3 mm) & $F_{\mathrm{int}}$ (C$^{17}$O)\tablenotemark{a}  & $F_{\mathrm{int}}$ (H$_2$CO)\tablenotemark{a} \\ 
    & (mJy beam$^{-1}$) & (mJy) & (Jy  km s$^{-1}$)) & (Jy km s$^{-1}$) \\
    \hline 
    \\[-.3cm]
    IRAS 04302+2247 &  \hspace{.8mm} 24.7 $\pm$ 0.1 & 165.9 $\pm$ 0.8 & 2.2 $\pm$ 0.2 & 3.5 $\pm$ 0.2\\
    L1489 IRS & \hspace{2mm} 2.8 $\pm$ 0.1 & \hspace{.8mm} 51.1 $\pm$ 1.1 & 2.9 $\pm$ 0.3 & 8.0 $\pm$ 0.5  \\   
    L1527 IRS & 102.0 $\pm$ 0.1 & 195.1 $\pm$ 0.4 & 1.9 $\pm$ 0.4 & 3.0 $\pm$  0.6 \\
    TMC1A & 125.8 $\pm$ 0.2 & 210.4 $\pm$ 0.4 & 4.1 $\pm$ 0.4 & 2.3 $\pm$ 0.2  \\
    TMC1-E &  \hspace{2mm} 9.2 $\pm$ 0.1 &  \hspace{.8mm} 10.3 $\pm$ 0.2 & \hspace{1mm} 2.0 $\pm$ 0.3\tablenotemark{b} & \hspace{1mm} 2.6 $\pm$ 0.2\tablenotemark{b}   \\
    TMC1-W &  \hspace{.8mm} 16.2 $\pm$ 0.1 &  \hspace{.8mm} 17.6 $\pm$ 0.2 & \hspace{1mm} 2.0 $\pm$ 0.3\tablenotemark{b} & \hspace{1mm} 2.6 $\pm$ 0.2\tablenotemark{b}   \\
    \hline
\end{tabular}
\tablecomments{The listed errors are statistical errors and do not include calibration uncertainties.}
\tablenotetext{a}{Integrated flux within a circular aperture with 6.0$^{\prime\prime}$ diameter.}
\vspace{-0.2cm}
\tablenotetext{b}{Flux for both sources together.}
\end{footnotesize}
\end{table*}


\section{Observations} \label{sec:Observations}

In order to study the temperature structure in young disks, a sample of five Class I protostars in Taurus was observed with ALMA: IRAS 04302+2247 (also known as the Butterfly star, hereafter IRAS 04302), L1489 IRS (hereafter L1489), L1527 IRS (hereafter L1527), TMC1 and TMC1A. All sources are known to have a disk and Keplerian rotation has been established (\citealt{Brinch2007,Tobin2012,Harsono2014}, van 't Hoff et al. in prep.). IRAS 04302 and L1527 are seen edge-on, which allows a direct view of the midplane, whereas L1489, TMC1 and TMC1A are moderately inclined $\sim$50--60$^\circ$. The source properties are listed in Table~\ref{tab:SourceOverview}.  

The observations were carried out on 2018 September 10 and 28, for a total on source time of 15 minutes per source (project code 2017.1.01413.S). The observations used 47 antennas sampling baselines between 15~m and 1.4~km. The correlator setup included a 2 GHz continuum band with 488 kHz (0.6 km s$^{-1}$) resolution centered at 240.0 GHz, and spectral windows targeting C$^{17}$O $2-1$, H$_2$CO $3_{1,2}-2_{1,1}$, HDO $3_{1,2}-2_{2,1}$ and several CH$_3$OH $5_K-4_K$ transitions. The spectral resolution was 122.1 kHz for CH$_3$OH and 61.0 kHz for the other lines, which corresponds to a velocity resolution of 0.15 and 0.08 km~s$^{-1}$, respectively. The properties of the targeted lines can be found in Table~\ref{tab:Lineparameters}.

Calibration was done using the ALMA Pipeline and version 5.4.0 of the Common Astronomy Software Applications (CASA; \citealt{McMullin2007}). The phase calibrator was J0438+3004, and the bandpass and flux calibrator was J0510+1800. In addition, we performed up to three rounds of phase-only self-calibration on the continuum data with solution intervals that spanned the entire scan length for the first round, as short as 60 s in the second round, and as short as 30 s in the third round. The obtained phase solutions were also applied to the line data. Imaging was done using \textit{tclean} in CASA version 5.6.1. The typical restoring beam size using Briggs weighting with a robust parameter of 0.5 is $0.42^{\prime\prime} \times 0.28^{\prime\prime}$ (59 $\times$ 39 AU) for the continuum images and $0.48^{\prime\prime} \times 0.31^{\prime\prime}$ (67 $\times$ 43 AU) for the line images. The continuum images have a rms of $\sim$0.07 mJy beam$^{-1}$, whereas the rms in the line images is $\sim$5 mJy beam$^{-1}$ channel$^{-1}$ for 0.08 km~s$^{-1}$ channels. The observed continuum and line flux densities are reported in Table~\ref{tab:Flux}. 

\begin{figure*}
\centering
\includegraphics[width=\textwidth,trim={0cm 9cm 0cm 1.4cm},clip]{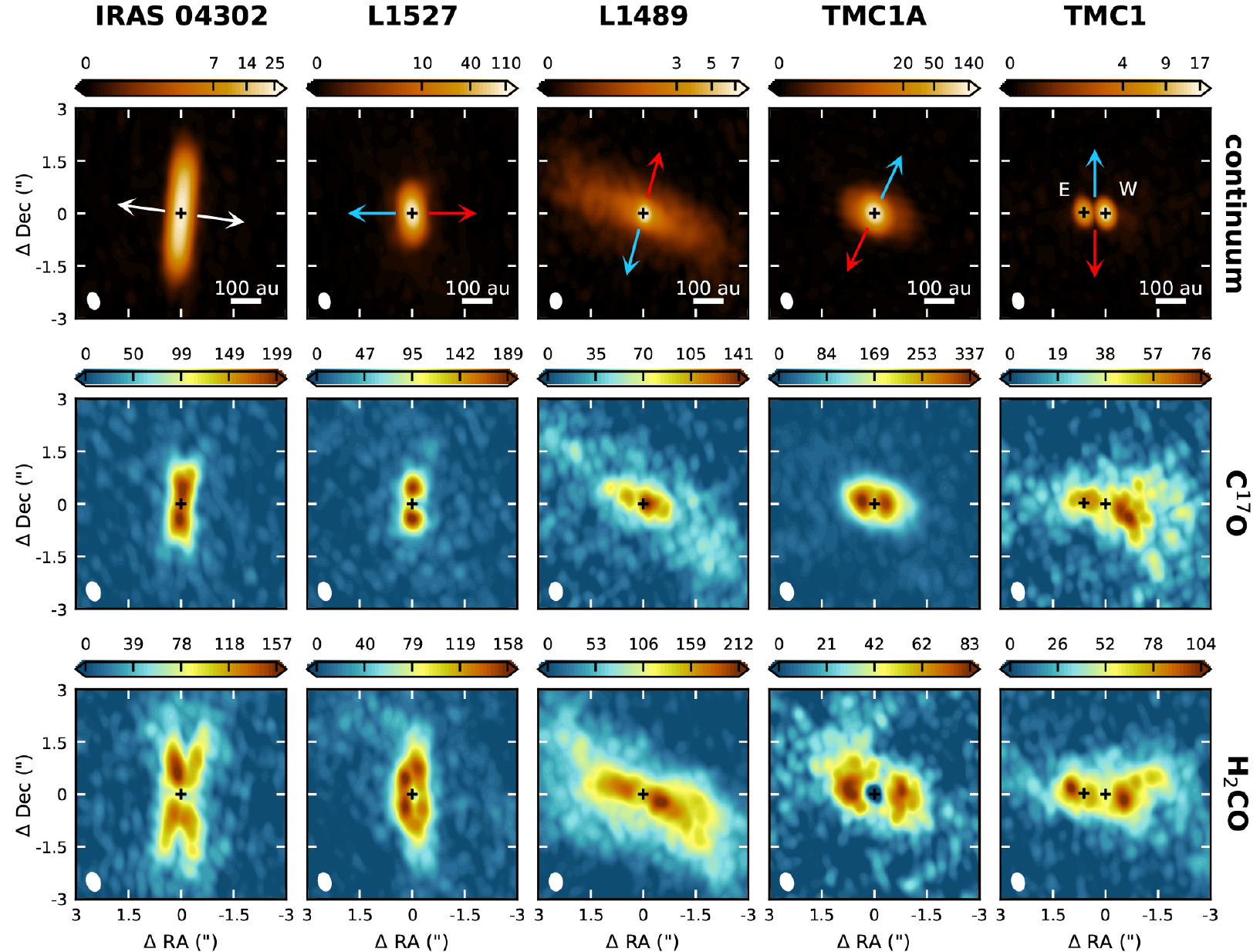}
\caption{Continuum images at 1.3 mm (\textit{top row}) and integrated intensity maps for the C$^{17}$O $2-1$ (\textit{middle row}) and H$_2$CO $3_{2,1}-2_{1,1}$ (\textit{bottom row}) transitions. The color scale is in mJy beam$^{-1}$ for the continuum images and in mJy beam$^{-1}$ km s$^{-1}$ for the line images. The positions of the continuum peaks are marked with black crosses, and the outflow directions are indicated by arrows in the continuum images. The beam is shown in the lower left corner of each panel.}
\label{fig:M0_overview}
\end{figure*}

\begin{figure*}
\centering
\includegraphics[width=0.8\textwidth,trim={0cm 0cm 2.8cm 0cm},clip]{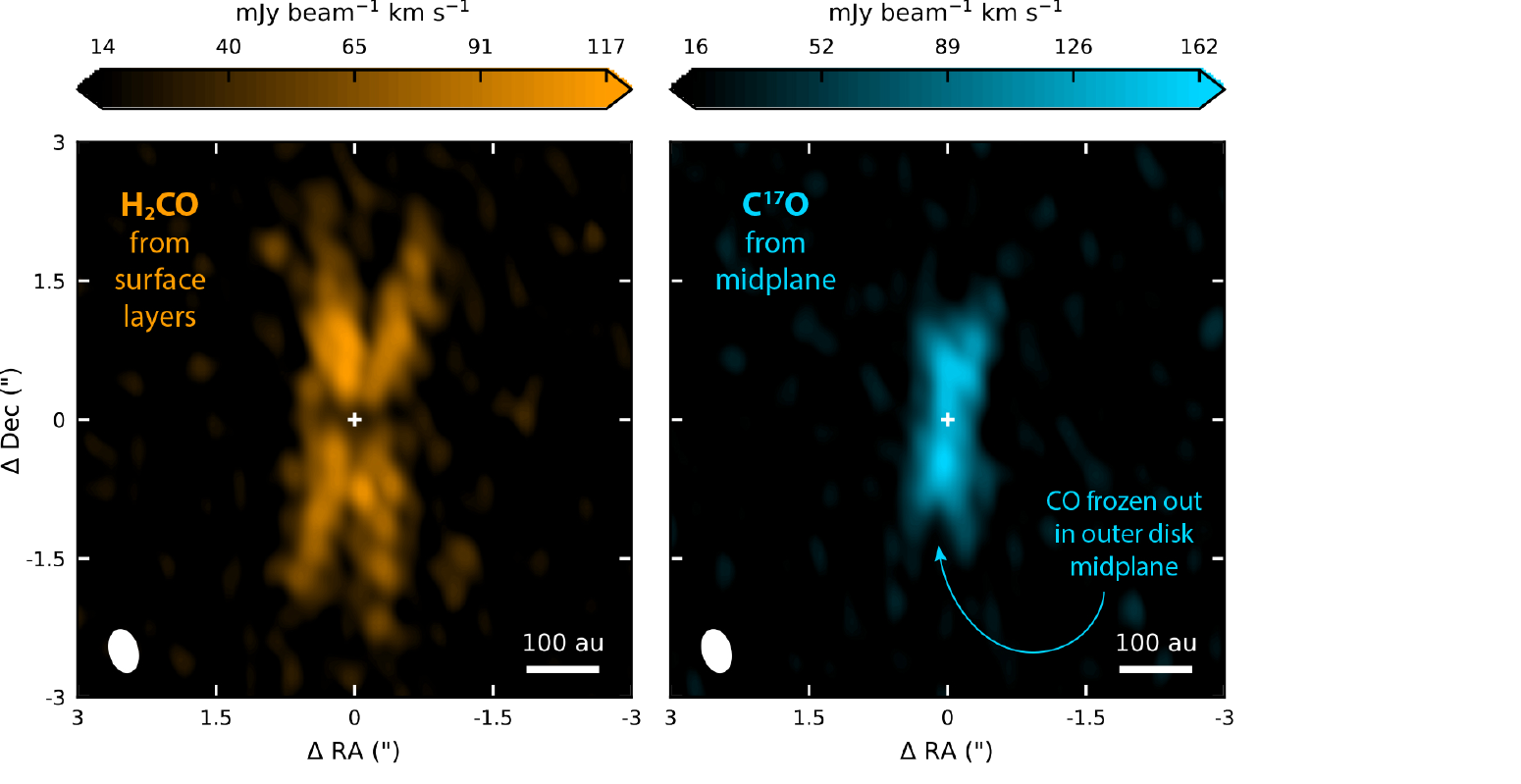}
\caption{Integrated intensity maps for the H$_2$CO $3_{2,1}-2_{1,1}$ (\textit{left}) and C$^{17}$O $2-1$ (\textit{right}) emission toward IRAS 04302. These images have slightly higher resolution than shown in Fig.~\ref{fig:M0_overview} ($0.45^{\prime\prime} \times 0.28^{\prime\prime}$) due to uniform weighting of the visibilities. The positions of the continuum peaks are marked with white crosses and the beam is shown in the lower left corner of each panel.}
\label{fig:IRAS04302}
\end{figure*}


\section{Results} \label{sec:Results}

\subsection{C$^{17}$O and H$_2$CO morphology}\label{sec:morphology}

\begin{figure}
\centering
\includegraphics[trim={0cm 1.6cm 0cm 1.5cm},clip]{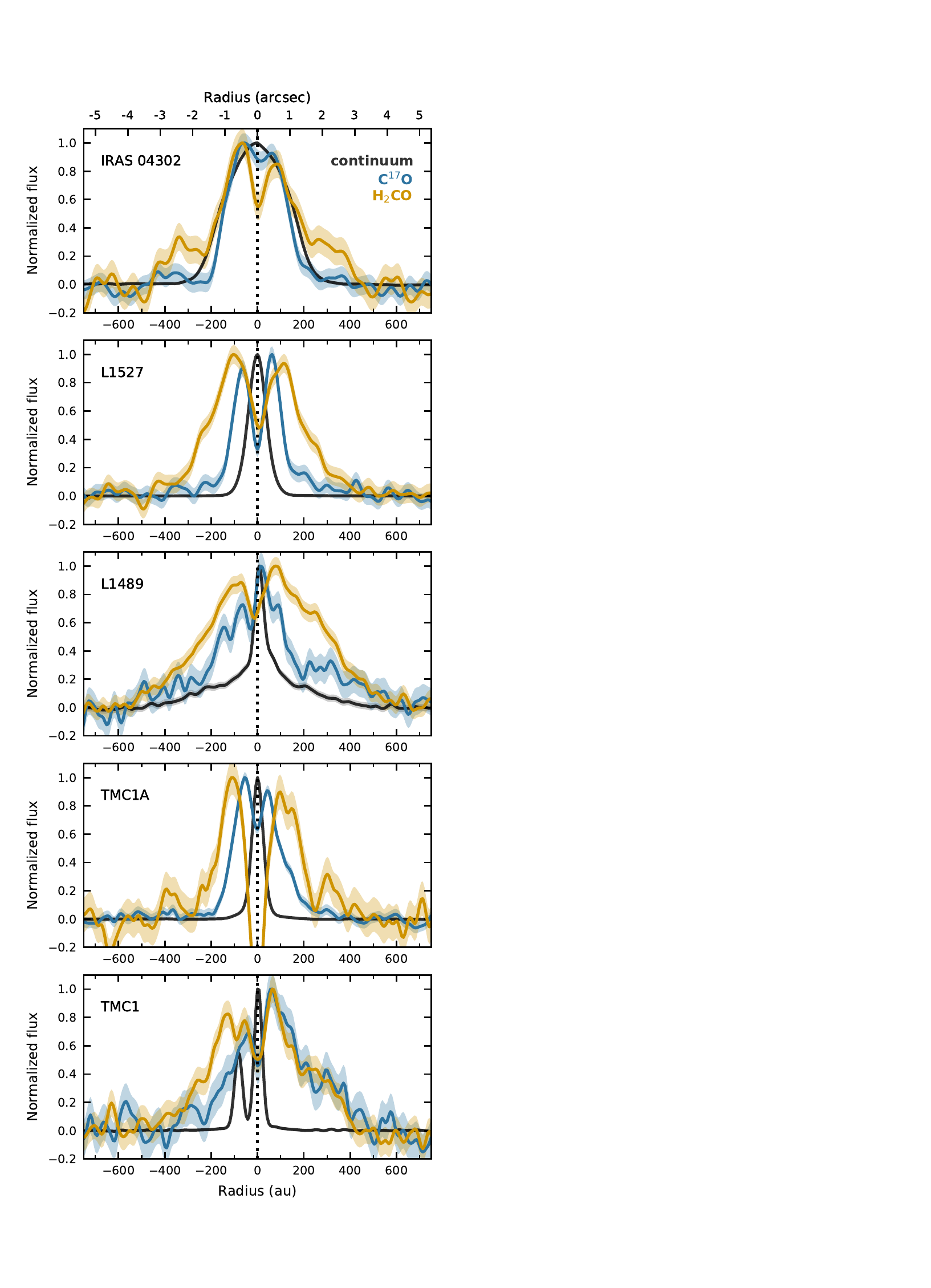}
\caption{Normalized radial cuts along the disk major axis for the 1.3 mm continuum flux (black) and the C$^{17}$O (blue) and H$_2$CO (orange) integrated intensities. The shaded area shows the 3$\sigma$ uncertainty.}
\label{fig:Radprofiles}
\end{figure}

Figure~\ref{fig:M0_overview} shows the 1.3 mm continuum images and integrated intensity (zeroth moment) maps for C$^{17}$O $2-1$ and H$_2$CO $3_{1,2}-2_{1,1}$ toward the five sources in our sample. The molecular emission toward IRAS 04302 is highlighted at slightly higher spatial resolution in Fig.~\ref{fig:IRAS04302}. Radial cuts along the major axis are presented in Fig.~\ref{fig:Radprofiles}. The continuum emission is elongated perpendicular to the outflow direction for all sources, consistent with a disk as observed before. TMC1 is for the first time resolved into a close binary ($\sim$85 AU separation). We will refer to the two sources as TMC1-E (east) and TMC1-W (west). 

Both C$^{17}$O and H$_2$CO are clearly detected toward all sources with a velocity gradient along the continuum structures (see Fig.~\ref{fig:M1_overview}). The velocity gradient suggests that the material in TMC1 is located in a circumbinary disk, but a detailed analysis is beyond the scope of this paper. For both molecules, integrated fluxes are similar (within a factor 2--3) in all sources (Table~\ref{tab:Flux}) and both lines have a comparable (factor 2--4) strength toward each source, with H$_2$CO brighter than C$^{17}$O, except for TMC1A. The H$_2$CO emission is generally more extended than the C$^{17}$O emission, both radially and vertically, except toward TMC1 and TMC1A where both molecules have the same spatial extent. This is not a signal-to-noise issue, as can be seen from the radial cuts along the major axis (Fig.~\ref{fig:Radprofiles}). 

The most striking feature in the integrated intensity maps is the V-shaped emission pattern of the H$_2$CO in the edge-on disk IRAS~04302 (see Fig.~\ref{fig:IRAS04302}), suggesting that the emission arises from the disk surface layers and not the midplane, in contrast to the C$^{17}$O emission. The H$_2$CO emission displays a ring-like structure toward L1527. Given that this disk is also viewed edge-on, this can be explained by emission originating in the disk surface layers, with the outer component along the midplane arising from the envelope. As we will show later in this section, the emission toward IRAS~04302 shows very little envelope contribution, which can explain the difference in morphology between these two sources. The C$^{17}$O emission peaks slightly offset ($\sim$60~au) from the L1527 continuum peak, probably due to the dust becoming optically thick in the inner $\sim$10~au as seen before for $^{13}$CO and C$^{18}$O \citep{vantHoff2018b}. The current resolution does not resolve the inner 10~au, hence the reduction in CO emission is more extended. In IRAS~04302, a similar offset of $\sim$60~au is found for both C$^{17}$O and H$_2$CO, suggesting there may be an unresolved optically thick dust component as well. 

Toward L1489, C$^{17}$O has a bright inner component ($\sim$200 au) and a weaker outer component that extends roughly as far as the H$_2$CO emission ($\sim$600 au). A similar structure was observed in C$^{18}$O by \citet{Sai2020}. The slight rise seen in C$^{18}$O emission around $\sim$300 au to the southwest of the continuum peak is also visible in the C$^{17}$O radial cut. Imaging the C$^{17}$O data at lower resolution makes this feature more clear in the integrated intensity map. In contrast, the H$_2$CO emission decreases in the inner $\sim$75 au, but beyond that it extends smoothly out to $\sim$600 au. The off-axis protrusions at the outer edge of the disk pointing to the northeast and to the southwest were also observed in C$^{18}$O and explained as streams of infalling material \citep{Yen2014}.  

The C$^{17}$O emission peaks slightly ($\sim$40--50 au) off-source toward TMC1A. \citet{Harsono2018} showed that $^{13}$CO and C$^{18}$O emission is absent in the inner $\sim$15~au due to the dust being optically thick. The resolution of the C$^{17}$O observations is not high enough to resolve this region, resulting only in a central decrease in emission instead of a gap. A clear gap is visible for H$_2$CO with the emission peaking $\sim$100--115 au off source. The central absorption falling below zero is an effect of resolved out large-scale emission. 

Finally, toward TMC1, H$_2$CO shows a dip at both continuum peaks, while the C$^{17}$O emission is not affected by the eastern continuum peak. As discussed for the other sources, this may be the result of optically thick dust in the inner disk. The protrusions seen on the west side in both C$^{17}$O and H$_2$CO are part of a larger arc-like structure that extends toward the southwest beyond the scale shown in the image. 

\begin{figure*}
\centering
\includegraphics[width=\textwidth]{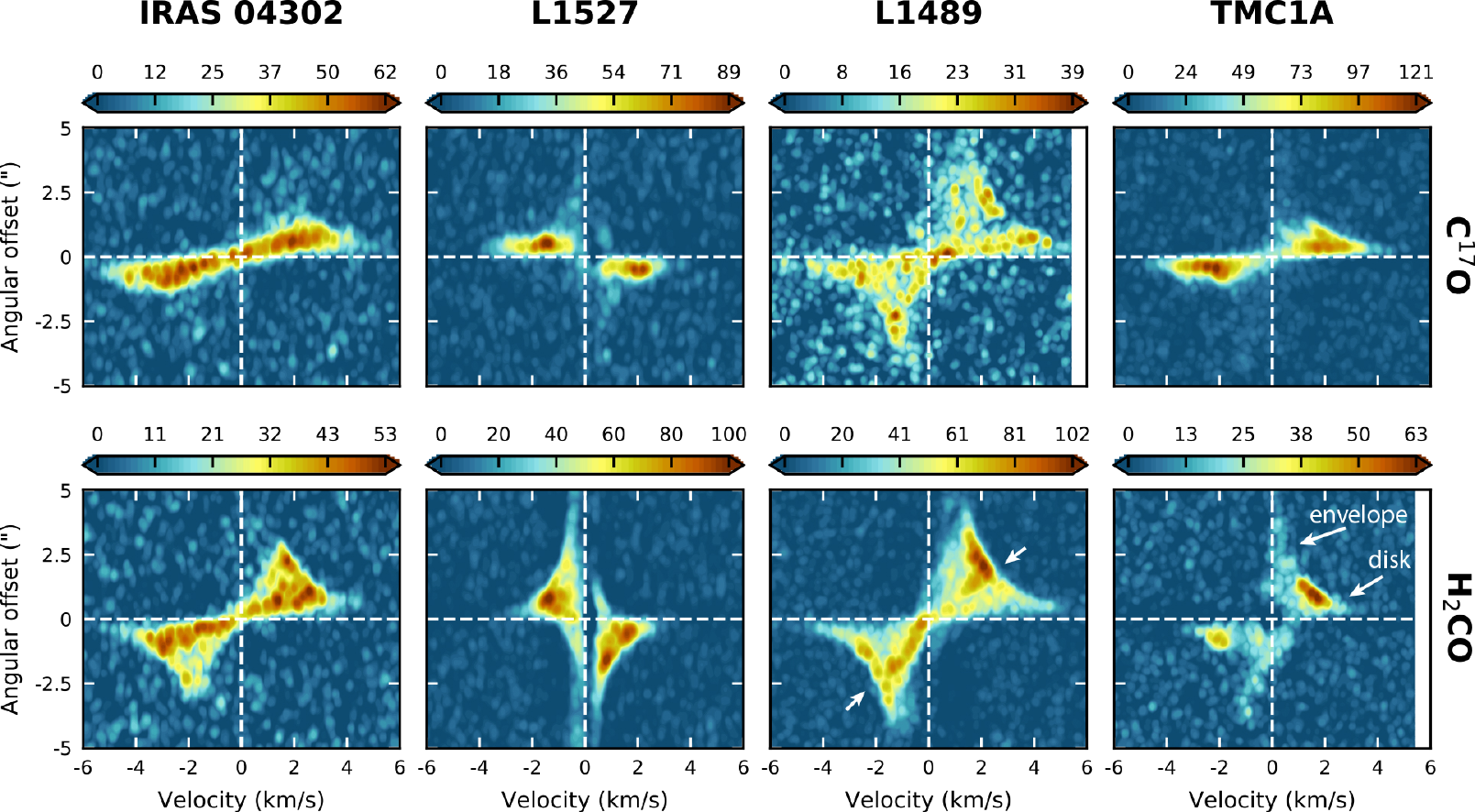}
\caption{Position-velocity diagrams for C$^{17}$O (\textit{top panels}) and H$_2$CO (\textit{bottom panels}) along the major axis of the disks in the single systems (listed above the rows). C$^{17}$O traces predominantly the disk, that is, high velocities at small angular offsets, whereas H$_2$CO generally has a larger envelope component, that is, low velocities at large angular offsets. The velocity is shifted such that 0 km s$^{-1}$ corresponds to the systemic velocity. The color scale is in mJy beam$^{-1}$. The white arrows in the L1489 H$_2$CO panel highlight the linear feature that is described in the text.}
\label{fig:PVdiagrams}
\end{figure*}

While it is tempting to ascribe all of the compact emission to the young disk, some of it may also come from the envelope and obscure the disk emission. To get a first impression whether the observed emission originates in the disk or in the envelope, position-velocity (\textit{pv}) diagrams are constructed along the disk major axis for the four single sources (Fig.~\ref{fig:PVdiagrams}). In these diagrams, disk emission is located at small angular offsets and high velocities, while envelope emission extends to larger offsets but has lower velocities. In all sources, C$^{17}$O traces predominantly the disk, with some envelope contribution, especially in L1527 and L1489. H$_2$CO emission also originates in the disk, but has a larger envelope component. An exception is IRAS 04302, which shows hardly any envelope contribution. These results for L1527 are in agreement with previous observations \citep{Sakai2014b}. In L1489, a bright linear feature is present for H$_2$CO extending from a velocity and angular offset of -2 km s$^{-1}$ and -2$^{\prime\prime}$, respectively, to offsets of 2 km s$^{-1}$ and 2$^{\prime\prime}$. This feature matches the shape of the SO \textit{pv}-diagram \citep{Yen2014}, which was interpreted by the authors as a ring between $\sim$250--390 au. While a brightness enhancement was also identified by \citet{Yen2014} in the C$^{18}$O emission (similar as seen here for H$_2$CO), the C$^{17}$O emission does not display such feature. 

Another way to determine the envelope contribution is from the visibility amplitudes. Although a quantitative limit on the envelope contribution to the line emission requires detailed modeling for the individual sources, which will be done in a subsequent paper, a first assessment can be made with more generic models containing either only a Keplerian disk or a disk embedded in an envelope (see Appendix~\ref{ap:Vismod}). For IRAS 04302, both the C$^{17}$O and H$_2$CO visibility amplitude profiles can be reproduced without an envelope. This suggests that there is very little envelope contribution for this source, consistent with the \textit{pv} diagrams. A disk is also sufficient to reproduce the visibility amplitudes at velocities $> |1|$ km s$^{-1}$ from the systemic velocity toward L1489, L1527 and TMC1A. For the low velocities a small envelope contribution is required. The line emission presented here is thus dominated by the disk.

Although both the C$^{17}$O and H$_2$CO emission originates predominantly from the disk, the C$^{17}$O emission extends to higher velocities than the H$_2$CO emission in IRAS~04302, L1527 and TMC1A. This is more easily visualized in the spectra presented in Fig.~\ref{fig:Spectra}. These spectra are extracted in a 6$^{\prime\prime}$ circular aperture and only include pixels with $> 3\sigma$ emission. While H$_2$CO is brighter at intermediate velocities than C$^{17}$O (even when correcting for differences in emitting area), it is not present at the highest velocities. H$_2$CO emission thus seems absent in the inner disk in these sources, which for TMC1A is also visible in the moment zero map (Fig.~\ref{fig:M0_overview}). However, in L1489, both molecules have similar maximum velocities. Toward TMC1 they extend to the same redshifted velocity, while C$^{17}$O emission is strongly decreased at blueshifted velocities as compared to the redshifted velocities. 

\subsection{C$^{17}$O and H$_2$CO column densities and abundances}

To compare the C$^{17}$O and H$_2$CO observations between the different sources more quantitatively, we calculate disk-averaged total column densities, $N_T$, assuming optically thin emission in local thermodynamic equilibrium (LTE), using
\begin{equation}\label{eq1}
 \frac{4\pi F\Delta v}{A_{ul}\Omega hcg_{\mathrm{up}}} = \frac{N_T}{Q(T_{\mathrm{rot}})}e^{-E_{\mathrm{up}}/kT_{\mathrm{rot}}}, 
\end{equation}
where $F\Delta v$ is the integrated flux density, $A_{ul}$ is the Einstein A coefficient, $\Omega$ is the solid angle subtended by the source, $E_{\mathrm{up}}$ and $g_{\mathrm{up}}$ are the upper level energy and degeneracy, respectively and $T_{\mathrm{rot}}$ is the rotational temperature. 

The integrated fluxes are measured over the dust emitting area (Table~\ref{tab:Coldens}). We note that this does not necessarily encompasses the total line flux, but it will allow for an abundance estimate as described below. A temperature of 30 K is adopted for C$^{17}$O and 100 K for H$_2$CO, as these are slightly above their freeze-out temperatures. The C$^{17}$O column density ranges between $\sim 2-20 \times 10^{15}$ cm$^{-2}$, with the lowest value toward L1489 and the highest value toward TMC1A. The H$_2$CO column density is about an order of magnitude lower with values between $\sim 4-18 \times 10^{14}$ cm$^{-2}$. The lowest value is found toward toward TMC1A and the highest value toward L1527. Changing the temperature for H$_2$CO to 30 K decreases the column densities by only a factor $\lesssim$3.
 
The H$_2$CO column density toward L1527 is a factor 3--6 higher than previously derived by \citet{Sakai2014b}, possibly because they integrated over different areas and velocity ranges for the envelope, disk and envelope-disk interface. Integrating the H$_2$CO emission over a circular aperture of 0.5$^{\prime\prime}$ and excluding the central $|\Delta v| \leq 1.0$ km s$^{-1}$ channels to limit the contribution from the envelope and resolved-out emission, results in a H$_2$CO column density of $9.7 \times 10^{13}$ cm$^{-2}$, only a factor 2--3 higher than found by \citet{Sakai2014b}. \citet{Pegues2020} found H$_2$CO column densities spanning three orders of magnitude ($\sim 5 \times 10^{11} - 5 \times 10^{14}$ cm$^{-2}$) for a sample of 13 Class II disks. The values derived here for Class I disks are thus similar to the high end ($\lesssim4$ times higher) of the values for Class II disks. 

An assessment of the molecular abundances can be made by estimating the H$_2$ column density from the continuum flux. First, we calculate the disk dust masses, $M_{\rm{dust}}$, from the integrated continuum fluxes, $F_\nu$, using 
\begin{equation} \label{eq2}
M_{\rm{dust}} = \frac{D^2 F_\nu}{\kappa_\nu B_\nu(T_{\rm{dust}})}, 
\end{equation}
where $D$ is the distance to the source, $\kappa_\nu$ is the dust opacity with the assumption of optically thin emission, and $B_\nu$ is the Planck function for a temperature $T_{\rm{dust}}$ \citep{Hildebrand1983}. Adopting a dust opacity of $\kappa_{1.3mm} = 2.25$ cm$^2$ g$^{-1}$, as used for Class II disks by e.g., \citet{Ansdell2016}, and a dust temperature of 30~K, similar to e.g., \citet{Tobin2015a} for embedded disks, results in disk dust mass between 3.7 $M_E$ for \mbox{TMC1-E} and 75 $M_E$ for TMC1A. Using the same dust opacity as for Class II disks is probably reasonable if grain growth starts early on in the disk-formation process. However, adopting $\kappa_{\rm{1.3mm}}$ = 0.899 cm$^2$ g$^{-1}$, as often done for protostellar disks and envelopes \citep[e.g.,][]{Jorgensen2007,Andersen2019,Tobin2020}, only affects the molecular abundances by a factor $\sim$2. Assuming a gas-to-dust ratio of 100 and using the size of the emitting region, these dust masses result in H$_2$ column densities of $2-90 \times 10^{23}$~cm$^{-2}$. 

The resulting C$^{17}$O and H$_2$CO abundances are listed in Table~\ref{tab:Coldens}. For C$^{17}$O, the abundances range between $1.2 \times 10^{-8}$ and $1.2 \times 10^{-7}$. Assuming a C$^{16}$O/C$^{17}$O ratio of 1792 (as in the interstellar medium; \citealt{Wilson1994}), a CO ISM abundance of 10$^{-4}$ with respect to H$_2$ corresponds to a C$^{17}$O abundance of $5.6 \times 10^{-8}$. The derived C$^{17}$O abundances are thus within a factor 5 of the ISM abundance, suggesting that no substantial processing has happened as observed for Class II disks where the CO abundance can be two orders of magnitude below the ISM value \citep[e.g.,][]{Favre2013}. These results are consistent with the results from \citet{Zhang2020} for three Class I disks in Taurus (including TMC1A), but not with the order of magnitude depletion found by \citet{Bergner2020} for two Class I disks in Serpens. For H$_2$CO, the abundance ranges between $\sim 3 \times 10^{-10}$--$\sim 8 \times 10^{-9}$ in the different sources, except for TMC1A where the abundance is 
$\sim 5 \times 10^{-11}$, probably due to the absence of emission in the inner region. Abundances around $10^{-10}$--$10^{-9}$ are consistent with chemical models for protoplanetary disks \citep[e.g.,][]{Willacy2009,Walsh2014}. However, H$_2$CO abundances derived for TW Hya and HD 163296 are 2--3 orders of magnitude lower, $8.9 \times 10^{-13}$ and $6.3 \times 10^{-12}$, respectively \citep{Carney2019}. 


\begin{table*}
\caption{Column densities and column density ratios.}
\label{tab:Coldens} 
\centering
\begin{footnotesize}
\begin{tabular}{l l c c c c c}
    \hline\hline
    \\[-.3cm]
     Source & Molecule & Area\tablenotemark{a} & $F_{\mathrm{int}}$\tablenotemark{b} & $N$\tablenotemark{c}  & $N$/$N$(H$_2$)\tablenotemark{d} & $N$/$N$(H$_2$CO)\tablenotemark{e}  \\ 
     & & ($^{\prime\prime}$ $\times$ $^{\prime\prime}$) & (Jy km s$^{-1}$) & (cm$^{-2}$) & & \\
    \hline 
    \\[-.3cm]
    IRAS 04302 & C$^{17}$O & 3.95 $\times$ 1.01 & \hspace{0.7mm} 1.4 $\pm$ 0.05 & $5.5 \pm 0.41 \times 10^{15}$ & \hspace{2.8mm} $3.8 \times 10^{-8}$ & \hspace{4mm} 46 \\
    & H$_2$CO & 3.95 $\times$ 1.01 &  \hspace{0.7mm} 1.5 $\pm$ 0.05 & $1.2 \pm 0.04 \times 10^{14}$ &  \hspace{3.5mm} $8.4 \times 10^{-10}$ &  \hspace{4mm} - \\
    & HDO & 0.50 $\times$ 0.50 & $< 4.5 \times 10^{-3}$ & \hspace{5mm} $< 7.4 \times 10^{13}$ & \hspace{.6mm} $< 5.3 \times 10^{-10}$ & \hspace{.4mm} $<$ 0.62\\
    & CH$_3$OH & 0.50 $\times$ 0.50 &  $< 6.9 \times 10^{-3}$ & \hspace{5mm} $< 7.3 \times 10^{14}$ & $< 5.2 \times 10^{-9}$ & $<$ 6.1\\
    
    L1489 & C$^{17}$O & 4.05 $\times$ 2.19 & \hspace{0.7mm} 1.5 $\pm$ 0.11 &  $ 2.3 \pm 0.40 \times 10^{15}$ & \hspace{2.8mm} $1.2 \times 10^{-7}$ & \hspace{3mm} 15  \\
    & H$_2$CO & 4.05 $\times$ 2.19 & \hspace{0.7mm} 4.2 $\pm$ 0.11 &  $1.5 \pm 0.04 \times 10^{14}$ & \hspace{3mm} $7.6 \times 10^{-9}$ & \hspace{4mm} - \\
    & HDO & 0.50 $\times$ 0.50 & $< 5.0 \times 10^{-3}$ & \hspace{5mm} $< 8.3 \times 10^{13}$ & $< 4.2 \times 10^{-9}$ & \hspace{2mm} $<$ 0.55\\
    & CH$_3$OH & 0.50 $\times$ 0.50 & $< 8.4 \times 10^{-3}$ & \hspace{5mm} $< 8.8 \times 10^{14}$ & $< 4.4 \times 10^{-8}$ & \hspace{.4mm} $<$ 5.9\\
    
    L1527 & C$^{17}$O & 1.34 $\times$ 0.77 & 0.54 $\pm$ 0.03 &  $ 7.7 \pm 0.96 \times 10^{15}$ & \hspace{2.5mm} $1.2 \times 10^{-8}$ & \hspace{3mm} 43 \\
    & H$_2$CO & 1.34 $\times$ 0.77 & 0.55 $\pm$ 0.03 &  $1.8 \pm 0.10 \times 10^{14}$& \hspace{4mm}$2.7 \times 10^{-10}$ & \hspace{4mm} - \\
    & HDO & 0.50 $\times$ 0.50 & $< 5.6 \times 10^{-3}$ & \hspace{5mm} $< 9.2 \times 10^{13}$ & \hspace{.6mm} $< 1.4 \times 10^{-10}$ & \hspace{2mm} $<$ 0.51\\
    & CH$_3$OH & 0.50 $\times$ 0.50 &  $< 7.9 \times 10^{-3}$ & \hspace{5mm} $< 8.3 \times 10^{14}$ & $< 1.3 \times 10^{-9}$ & \hspace{.4mm} $<$ 4.6 \\
    
    TMC1A & C$^{17}$O & 0.93 $\times$ 0.88 & \hspace{0.7mm} 1.1 $\pm$ 0.02 &  $2.0 \pm 0.08 \times 10^{16}$ & \hspace{3mm} $2.3 \times 10^{-8}$ & \hspace{6mm} 488 \\
    & H$_2$CO & 0.93 $\times$ 0.88 & 0.10 $\pm$ 0.02 &  $4.1 \pm 0.82 \times 10^{13}$ &  \hspace{4mm}$4.6 \times 10^{-11}$  & \hspace{4mm} -\\
    & HDO & 0.50 $\times$ 0.50 & $< 5.0 \times 10^{-3}$ & \hspace{5mm} $< 8.3 \times 10^{13}$ & \hspace{.6mm} $< 9.3 \times 10^{-11}$ &  \hspace{.4mm} $<$ 2.0 \\
    & CH$_3$OH & 0.50 $\times$ 0.50 & $< 7.7 \times 10^{-3}$ & \hspace{5mm} $< 8.1 \times 10^{14}$ & \hspace{.6mm} $< 9.1 \times 10^{-10}$ & \hspace{.4mm} $<$ 18\\
    
    TMC1-E & C$^{17}$O & 0.71 $\times$ 0.54 & 0.10 $\pm$ 0.01 &  $3.6 \pm 0.85 \times 10^{15}$ & \hspace{2.8mm} $3.9 \times 10^{-8}$ & \hspace{3mm} 33\\
    & H$_2$CO & 0.71 $\times$ 0.54 & 0.12 $\pm$ 0.01 &  $1.1 \pm 0.09 \times 10^{14}$ & \hspace{3mm}$1.2 \times 10^{-9}$ & \hspace{4mm} - \\
    & HDO & 0.50 $\times$ 0.50 & $< 5.0 \times 10^{-3}$ & \hspace{5mm} $< 8.3 \times 10^{13}$ & \hspace{.6mm} $< 8.9 \times 10^{-10}$ & \hspace{.4mm} $<$ 0.75\\
    & CH$_3$OH & 0.50 $\times$ 0.50 & $< 7.7 \times 10^{-3}$ & \hspace{5mm} $< 8.1 \times 10^{14}$ & $< 8.7 \times 10^{-9}$ &  \hspace{.4mm} $<$ 7.4\\
    
    TMC1-W & C$^{17}$O & 0.81 $\times$ 0.63 & 0.12 $\pm$ 0.01 &  $ 3.3 \pm 0.65 \times 10^{15}$ & \hspace{2.8mm} $2.8 \times 10^{-8}$ & \hspace{3mm} 35\\
    & H$_2$CO & 0.81 $\times$ 0.63 & 0.15 $\pm$ 0.01 &  $9.5 \pm 0.66 \times 10^{13}$ & \hspace{3mm}$8.0 \times 10^{-9}$ & \hspace{4mm} - \\
    & HDO & 0.50 $\times$ 0.50 & $< 5.0 \times 10^{-3}$ & \hspace{5mm} $< 8.3 \times 10^{13}$ & \hspace{.6mm} $< 6.9 \times 10^{-10}$ & \hspace{.4mm} $<$ 0.87\\
    & CH$_3$OH & 0.50 $\times$ 0.50 & $< 7.7 \times 10^{-3}$ & \hspace{5mm} $< 8.1 \times 10^{14}$ & $< 6.8 \times 10^{-9}$ &  \hspace{.4mm} $<$ 8.5\\
    \hline
\end{tabular}
\begin{flushleft}
\vspace{-0.3cm} 
\tablenotetext{a}{Area over which the flux is extracted.}
\vspace{-0.2cm} 
\tablenotetext{b}{Integrated flux. For HDO and CH$_3$OH this is the 3$\sigma$ upper limit to the integrated flux.}
\vspace{-0.2cm} 
\tablenotetext{c}{Column density.}
\vspace{-0.2cm} 
\tablenotetext{d}{Column density with respect to H$_2$, where the H$_2$ column density estimated from the continuum flux and assuming a gas-to-dust ratio of 100.}
\vspace{-0.2cm} 
\tablenotetext{e}{Column density with respect to H$_2$CO.}
\end{flushleft}
\end{footnotesize}
\end{table*}

Caveats in determining these abundances are the assumption that the continuum and line emission is optically thin. As discussed in Sect.~\ref{sec:morphology}, there is likely an optially thick dust component which would result in underestimates of the dust masses and overestimates of the abundances. On the other hand, optically thick dust hides molecular line emission originating below its $\tau = 1$ surface, which leads to underestimates of the abundances. Based on the results from \citet{Zhang2020}, C$^{17}$O may be optically thick in Class I disks. This would also result in underestimating the abundances. Scaling the dust temperature used in Eq.~\ref{eq2} with luminosity as done by \citet{Tobin2020} for embedded disks in Orion, results in dust masses lower by a factor $\sim$2, and therefore slightly higher abundances. Moreover, the integrated line flux is assumed to originate solely in the disk, but as shown in Fig.~\ref{fig:PVdiagrams}, there can be envelope emission present. Finally, the H$_2$CO emission originates in the disk surface layers, which means the abundances are higher than derived here assuming emission originating throughout the disk. To take all these effects in to account, source specific models are required. 

\subsection{HDO and CH$_3$OH upper limits}

Water and methanol form on ice-covered dust grains and thermally desorb into the gas phase at temperatures $\sim$100--150 K. These molecules are thus expected to trace the hot region inside the water snowline. The observations cover one HDO (deuterated water) transition ($3_{1,2}-2_{2,1}$) with an upper level energy of 168 K, and 16 transitions in the CH$_3$OH $J = 5_k - 4_k$ branch with upper level energies ranging between 34 and 131~K. None of these lines are detected in any of the disks. 

To compare these non-detections to observations in other systems, a 3$\sigma$ upper limit is calculated for the disk-averaged total column density by substituting  
\begin{equation} \label{eq3}
3\sigma = 3 \times1.1 \sqrt{\delta v\Delta V} \times \mathrm{rms}, 
\end{equation}
for the integrated flux density, $F\Delta v$ in eq.~\ref{eq1}. Here $\delta v$ is the velocity resolution, and $\Delta V$ is the line width expected based on other line detections. The factor 1.1 takes a 10\% calibration uncertainty in account. Assuming the water and methanol emission arises from the innermost part of the disk, the rms is calculated from the base line of the spectrum integrated over a central 0.5$^{\prime\prime}$ diameter aperture ($\sim$ one beam) and amounts to $\sim$2.7 mJy for HDO and $\sim$3.0 mJy for CH$_3$OH. A line width of 4 km s$^{-1}$ and a rotational temperature of 100 K are adopted.

A 3$\sigma$ column density upper limit of $\sim$ $8\times10^{13}$ cm$^{-2}$ is then found for HDO. This is 1--2 orders of magnitude below the column densities derived for the Class 0 sources NGC1333 IRAS2A, NGC1333 IRAS4A-NW and NGC1333 IRAS4B ($\sim 10^{15} - 10^{16}$ cm$^{-2}$; \citealt{Persson2014}) and more than 3 orders of magnitude lower than toward the Class 0 source IRAS 16293A ($\sim 5 \times 10^{17}$ cm$^{-2}$; \citealt{Persson2013}). Taking into account the larger beam size of the earlier observations ($\sim1^{\prime\prime}$) lowers the here derived column density by only a factor $\sim$4. Furthermore, \citet{Taquet2013b} showed that the HDO observations toward NGC1333 IRAS2A and NGC1333 IRAS4A are consistent with column densities up to $10^{19}$ and $10^{18}$ cm$^{-2}$, respectively, using a grid of non-LTE large velocity gradient (LVG) radiative transfer models. 

For CH$_3$OH, the $5_{0,5}-4_{0,4}$ (A) transition provides the most stringent upper limit of $\sim$ $8\times10^{14}$ cm$^{-2}$. This upper limit is orders of magnitude lower than the column density toward the Class 0 source IRAS 16293 ($2 \times 10^{19}$ cm$^{-2}$ within a 70 au beam; \citealt{Jorgensen2016}) and the young disk around the outbursting star V883 Ori (disk-averaged column density of $\sim$ $1.0 \times 10^{17}$ cm$^{-2}$; \citealt{vantHoff2018c}). A similarly low upper limit ($5\times10^{14}$ cm$^{-2}$) was found for a sample of 12 Class I disks in Ophiuchus \citep{ArturdelaVillarmois2019}. However, this upper limit is not stringent enough to constrain the column down to the value observed in the TW Hya protoplanetary disk (peak column density of $3-6 \times 10^{12}$ cm$^{-2}$; \citealt{Walsh2016}) or the upper limit in the Herbig Ae disk HD 163296 (disk-averaged upper limit of $5\times 10^{11}$ cm$^{-2}$; \citealt{Carney2019}). 

\begin{figure*}
\centering
\includegraphics[width=\textwidth,trim={0cm 11.8cm 0cm 0.6cm},clip]{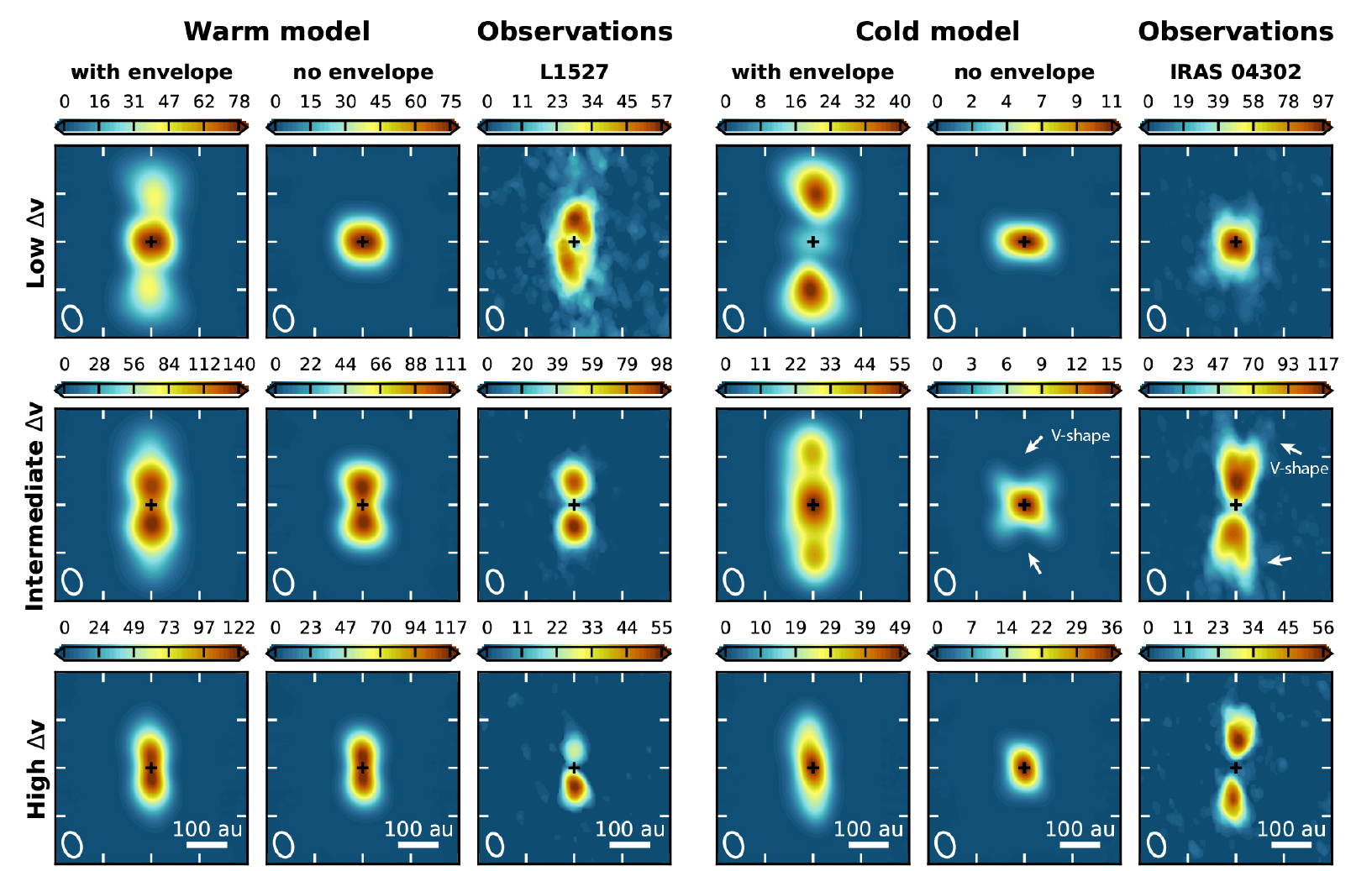}
\caption{Integrated intensity (moment zero) maps of the low-velocity (\textit{top row}), intermediate-velocity (\textit{middle row}), and high-velocity (\textit{bottom row}) C$^{17}$O emission in the warm (\textit{first and second column}) and cold edge-on disk models (\textit{fourth and fifth column}), as well as for the observations toward L1527 (\textit{third column}) and IRAS~04302 (\textit{sixth column}). The models contain either a disk and envelope (\textit{first and fourth column}) or only a disk (\textit{second and fifth column}). For the models, low velocities range from -1.0 to 1.0 km~s$^{-1}$, for intermediate velocities $|\Delta v|$ = 1.0-2.0 km~s$^{-1}$ and for high velocities $|\Delta v|$ = 2.0-4.0 km~s$^{-1}$ with respect to the source velocity. For IRAS 04302 (L1527), low velocities range from -1.19 to 1.09 (-1.19 to 1.25) km~s$^{-1}$, intermediate velocities range from -3.56 to -1.19 (-2.42 to -1.19) km~s$^{-1}$ and from 1.09 to 2.97 (1.25 to 2.39) km~s$^{-1}$, and high velocities range from -3.56 to -5.28 (-2.42 to -3.97) km~s$^{-1}$ and from 2.97 to 4.67 (2.39 to 3.13) km~s$^{-1}$ with respect to the source velocity. Only pixels with $>3\sigma$ emission are included. The color scale is in mJy beam$^{-1}$ km s$^{-1}$. The source position is marked with a black cross and the beam is shown in the lower left corner of the panels. A 100 au scalebar is present in the \textit{bottom panels.} The V-shaped emission pattern that is visible at intermediate velocities in the cold model and the IRAS 04302 observations is indicated by white arrows.}
\label{fig:M0_edgeondisks}
\end{figure*}

\begin{figure*}
\centering
\includegraphics[trim={0cm 9cm 0cm 1.5cm},clip]{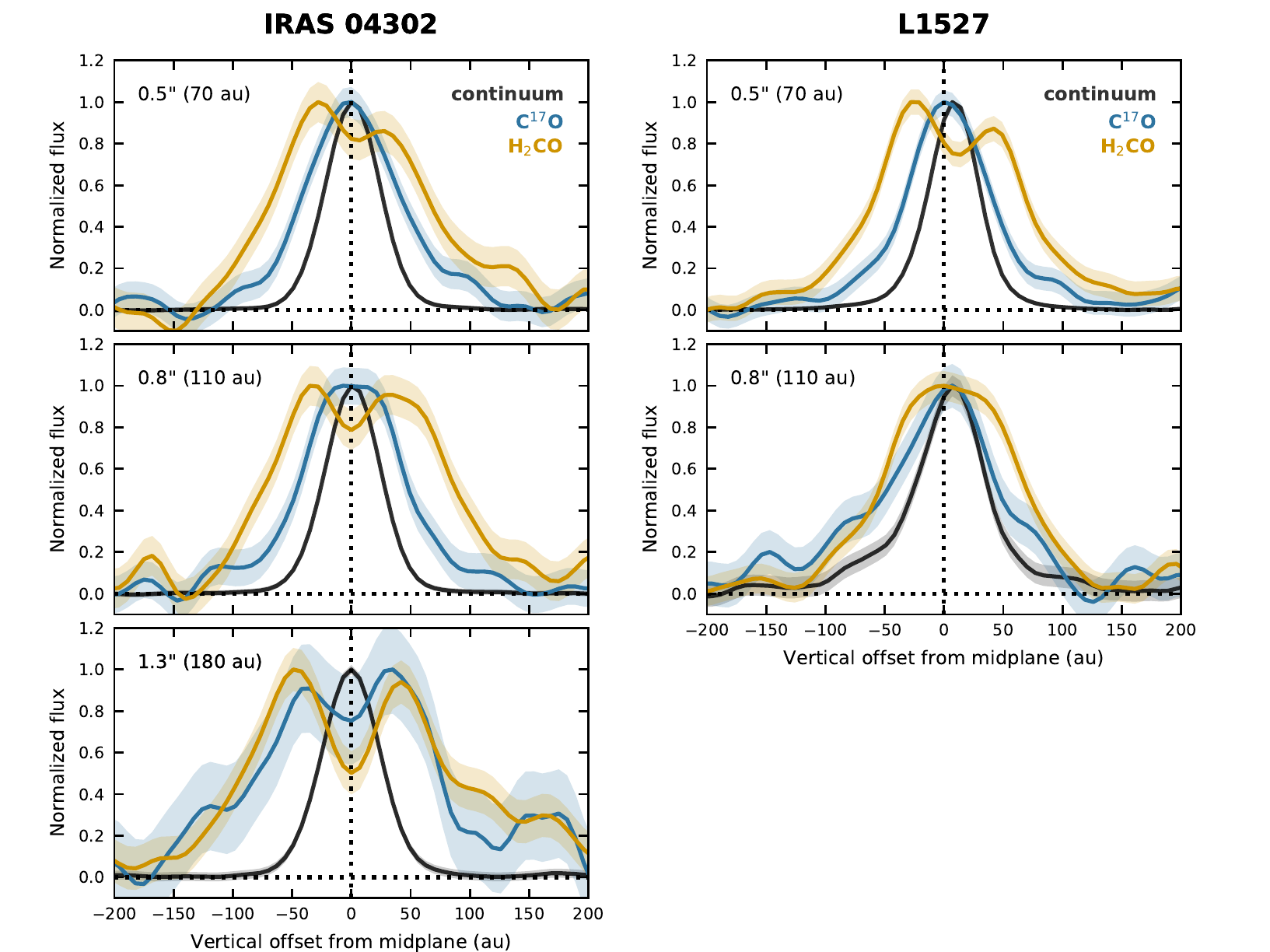}
\caption{Vertical cuts through the edge-on disks IRAS 04302 (\textit{left panels}) and L1527 (\textit{right panels}) at 0.5$^{\prime\prime}$ (\textit{top panels}), 0.8$^{\prime\prime}$ (\textit{middle panels}) and 1.3$^{\prime\prime}$ (\textit{bottom panel}) north of the continuum peak.  The 1.3 mm continuum is shown in black and the integrated intensity for C$^{17}$O $J=2-1$ and H$_2$CO 3$_{1,2}$-2$_{1,1}$ in blue and orange, resp. The shaded area shows the 3$\sigma$ uncertainty. The largest offset is not shown for L1527 because the continuum and C$^{17}$O emission reaches the noise limit. The H$_2$CO emission is single peaked at $\sim$10 au.}
\label{fig:Verticalprofiles}
\end{figure*}

For a better comparison with other sources, column density ratios are calculated with respect to H$_2$ and H$_2$CO, and are reported in Table~\ref{tab:Coldens}. Using the H$_2$ column density derived from the continuum flux, upper limits of $\sim 1-40 \times 10^{-10}$ are found for the HDO abundance. CH$_3$OH upper limits range between $1-40 \times 10^{-9}$. This is orders of magnitude lower than what is expected from ice observations ($10^{-6}-10^{-5}$; \citealt{Boogert2015}), and thus from thermal desorption, as observed in IRAS 16293 ($\lesssim 3 \times 10^{-6}$; \citealt{Jorgensen2016}) and V883-Ori ($\sim 4 \times 10^{-7}$; \citealt{vantHoff2018c}). Abundances for non-thermally desorbed CH$_3$OH in TW Hya are estimated to be $\sim 10^{-12}-10^{-11}$ \citep{Walsh2016}. \citet{Sakai2014b} detected faint CH$_3$OH emission (from different transitions than targeted here) toward L1527, with a CH$_3$OH/H$_2$CO ratio between 0.6 and 5.1. Our upper limit of 4.6 for L1527 is consistent with these values. CH$_3$OH/H$_2$CO ratios of 1.3 and $<$ 0.2 were derived for TW Hya and HD 163296, respectively, but our CH$_3$OH upper limit is not stringent enough to make a meaningful comparison. An assumption here is that the emitting regions of CH$_3$OH and H$_2$CO are co-spatial. As noted in Sect.~\ref{sec:morphology}, H$_2$CO seems absent in the inner disk where CH$_2$OH is expected.  

\begin{figure*}
\centering
\includegraphics[trim={0cm 16.5cm 0cm 1cm},clip]{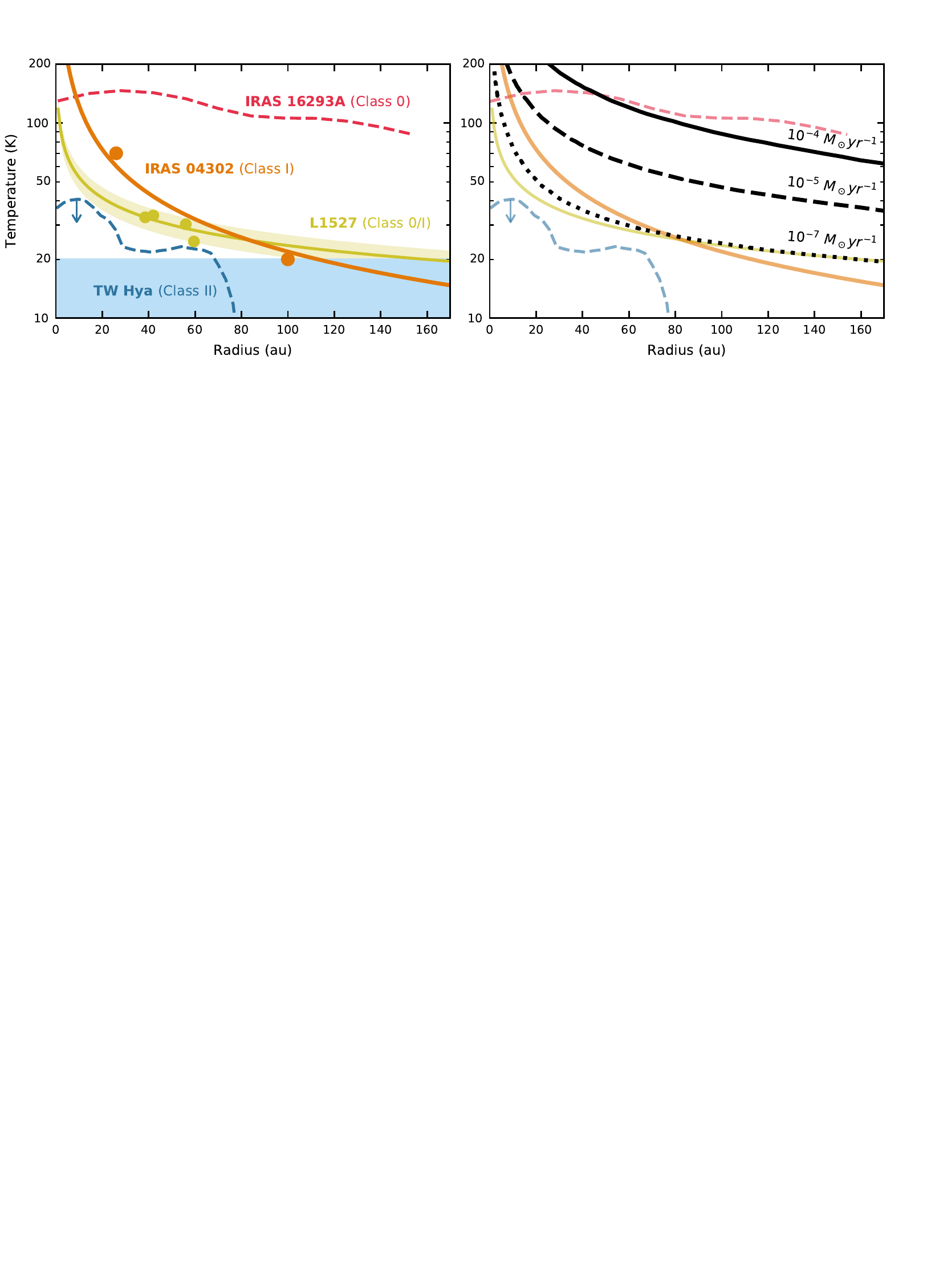}
\caption{\textit{Left panel:} Radial midplane temperature profile for IRAS 04302 inferred from the CO and H$_2$CO snowline estimates (orange circles). The solid orange line is a power law of the shape $T \propto R^{-0.75}$. For comparison, the temperature measurements for L1527 from  $^{13}$CO and C$^{18}$O emission (yellow circles) and a power law temperature profile with $T \propto R^{-0.35}$ (yellow line, with 1$\sigma$ uncertainty) are shown \citep{vantHoff2018b}, as well as the temperature profile derived for the disk-like structure in the Class~0 source IRAS 16293A (dashed red line; \citealt{vantHoff2020}), and the temperature profile for the Class II disk TW Hya (dashed blue line; \citealt{Schwarz2016}). The TW Hya temperature profile traces a warmer layer above the midplane and the midplane CO snowline is located around $\sim$20 AU \citep[e.g.,][]{vantHoff2017,Zhang2017}. The blue shaded area denotes the temperatures at which CO is frozen out. \textit{Right panel:} Temperature profiles from the \textit{left panel} overlaid with temperature profiles from embedded disk models from \citet{Harsono2015}. All three models have a stellar luminosity of 1~$L_\sun$, an envelope mass of 1~$M_\sun$, a disk mass of 0.05~$M_\sun$ and a disk radius of 200 au, but different accretion rates of $10^{-4} M_\sun$ yr$^{-1}$ (solid black line), $10^{-5} M_\sun$ yr$^{-1}$ (dashed black line) and $10^{-7} M_\sun$ yr$^{-1}$ (dotted black line) and therefore different total luminosities. }
\label{fig:PowerlawT}
\end{figure*}


\section{Analysis} \label{sec:Analysis}

\subsection{Temperature structure in the edge-on disks}

For (near) edge-on disks, CO freeze-out should be readily observable as CO emission will be missing from the outer disk midplane \citep{Dutrey2017,vantHoff2018b}. \citet{vantHoff2018b} studied the effect of CO freeze-out on the optically thick $^{13}$CO and C$^{18}$O emission in L1527. The less-abundant C$^{17}$O is expected to be optically thin and traces mainly the disk. Here we employ the models from \citet{vantHoff2018b} to predict the C$^{17}$O emission pattern for varying degrees of CO freeze-out (see Fig.~\ref{fig:diskmodels}): a `warm' model (no CO freeze-out), an `intermediate' model (CO freeze-out in the outer disk midplane) and a `cold' model (CO freeze-out in most of the disk, except the inner part and surface layers). Briefly, in these models gaseous CO is present at a constant abundance of 10$^{-4}$ with respect to H$_2$ in the regions in the disk where $T >$ 20 K and in the envelope. For the warm model, the L1527 temperature structure from \citet{Tobin2013} is adopted, and for the intermediate and cold models the temperature is reduced by 40\% and 60\%, respectively. There is no CO freeze out in the 125 au disk in the warm model, while the intermediate and cold models have the CO snowline at 71 and 23 au, respectively. Synthetic images cubes are generated using the radiative transfer code LIME \citep{Brinch2010}, making use of the C$^{17}$O LAMDA file \citep{Schoier2005} for the LTE calculation, and are convolved with the observed beam size. 

Figure~\ref{fig:M0_edgeondisks} shows moment zero maps integrated over the low, intermediate and high velocities for the warm and cold edge-on disk model. Models with and without an envelope are presented. The difference between the warm and cold model is most clearly distinguishable at intermediate velocities (Fig.~\ref{fig:M0_edgeondisks}, middle row). In the absence of an envelope, the emission becomes V-shaped in the cold model, tracing the warm surface layers where CO is not frozen out. This V-shape is not visible when there is a significant envelope contribution. The cold model differs from the warm model in that the envelope emission becomes comparable in strength to the disk emission when CO is frozen out in most of the disk. In the warm case, the disk emission dominates over the envelope emission. At low velocities (Fig.~\ref{fig:M0_edgeondisks}, top row), the difference between a warm and cold disk can be distinguished as well when an envelope is present, although in practice this will be much harder due to resolved out emission at these central velocities. Without an envelope, the low velocity emission originates near the source center due to the rotation, and the models are indistinguishable, except for differences in the flux. Due to the rotation, the emission at these velocities gets projected along the minor axis of the disk (that is, east-west). At the highest velocities (Fig.~\ref{fig:M0_edgeondisks}, top row), the emission originates in the inner disk, north and south of the source. If CO is absent in the midplane, very high angular resolution is be required to observe this directly through a V-shaped pattern. 

C$^{17}$O moment zero maps integrated over different velocity intervals for IRAS 04302 and L1527 are presented in Fig.~\ref{fig:M0_edgeondisks}. The observations show no sign of CO freeze-out in L1527 and resemble the warm model (most clearly seen at intermediate velocities), consistent with previous results for C$^{18}$O and $^{13}$CO \citep{vantHoff2018b}. IRAS 04302 on the other hand displays a distinct V-shaped pattern at intermediate velocities, suggesting that CO is frozen out in the outer part of this much larger disk ($\sim$250 au, compared to 75--125 au for L1527; \citealt{Aso2017,Tobin2013,Sheehan2017}). 

The vertical distribution of the emission in both disks is highlighted in Fig.~\ref{fig:Verticalprofiles} with vertical cuts at different radii. In L1527, the C$^{17}$O emission peaks at the midplane throughout the disk, while for IRAS~04302 the peaks shift to layers higher up in the disk for radii $\gtrsim$110 au. A first estimate of the CO snowline location can be made based on the location of the V-shape. In the cold model, the CO snowline is located at 23 au, but due to the size of the beam, the base of the V-shape and the first occurrence of a double peak in the vertical cuts is at $\sim$55 au. In IRAS 04302, the V-shape begins at a radius of $\sim$130 au, so the CO snowline location is then estimated to be around $\sim$100 au. 

\begin{figure}
\centering
\includegraphics[width=0.5\textwidth,trim={0cm 11cm 7.5cm 1.5cm},clip]{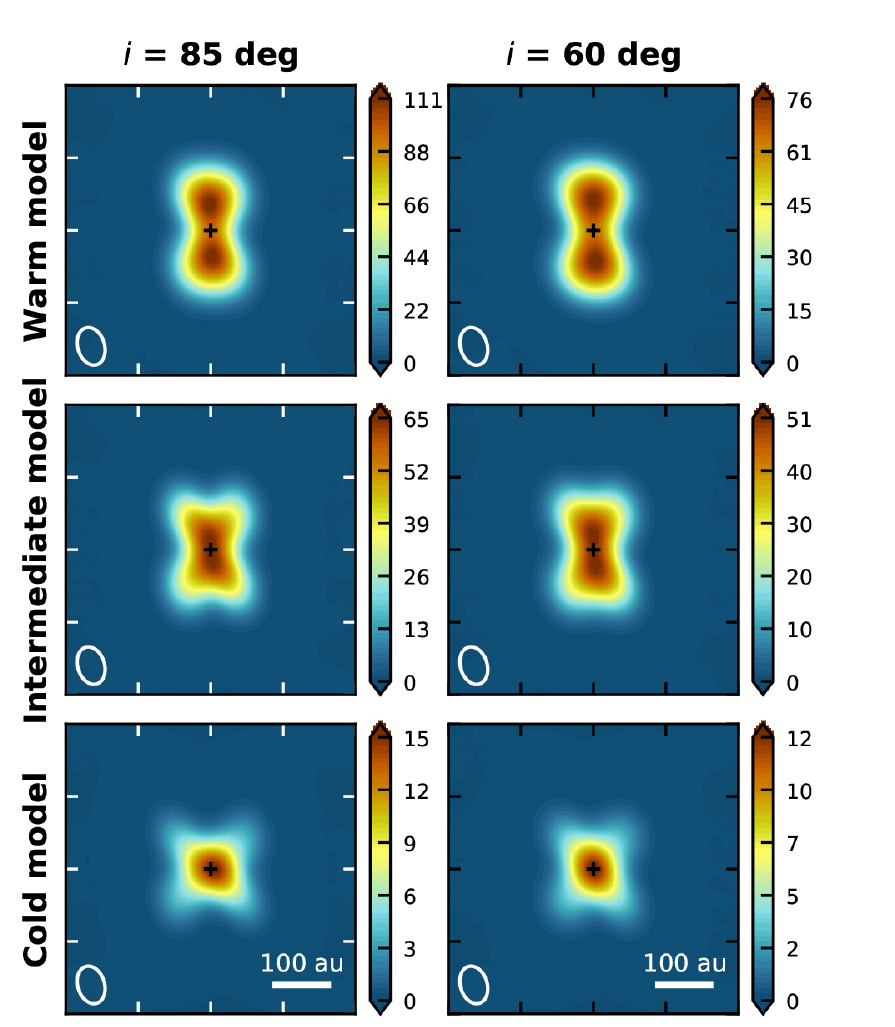}
\caption{Integrated intensity (moment zero) maps of the intermediate-velocity C$^{17}$O $J=2-1$ emission in the warm (\textit{top row}), intermediate (\textit{middle row}) and cold disk model (\textit{bottom row}). The \textit{left column} shows a near edge-on disk ($i=85^\circ$) as in Fig.~\ref{fig:M0_edgeondisks}, and the \textit{right column} shows a less-inclined disk ($i=85^\circ$). The velocity range $\Delta v$ is 1.0--1.9 km~s$^{-1}$ for $i=85^\circ$ and 1.3--1.8 km~s$^{-1}$ for $i=60^\circ$. The color scale is in mJy beam$^{-1}$ km s$^{-1}$. The source position is marked with a black cross and the beam is shown in the lower left corner of the panels. A 100 au scalebar is present in the bottom panels.}
\label{fig:M0_incl-effect}
\end{figure}

\begin{figure}
\centering
\includegraphics[width=0.5\textwidth,trim={0cm 11cm 7.5cm 1.5cm},clip]{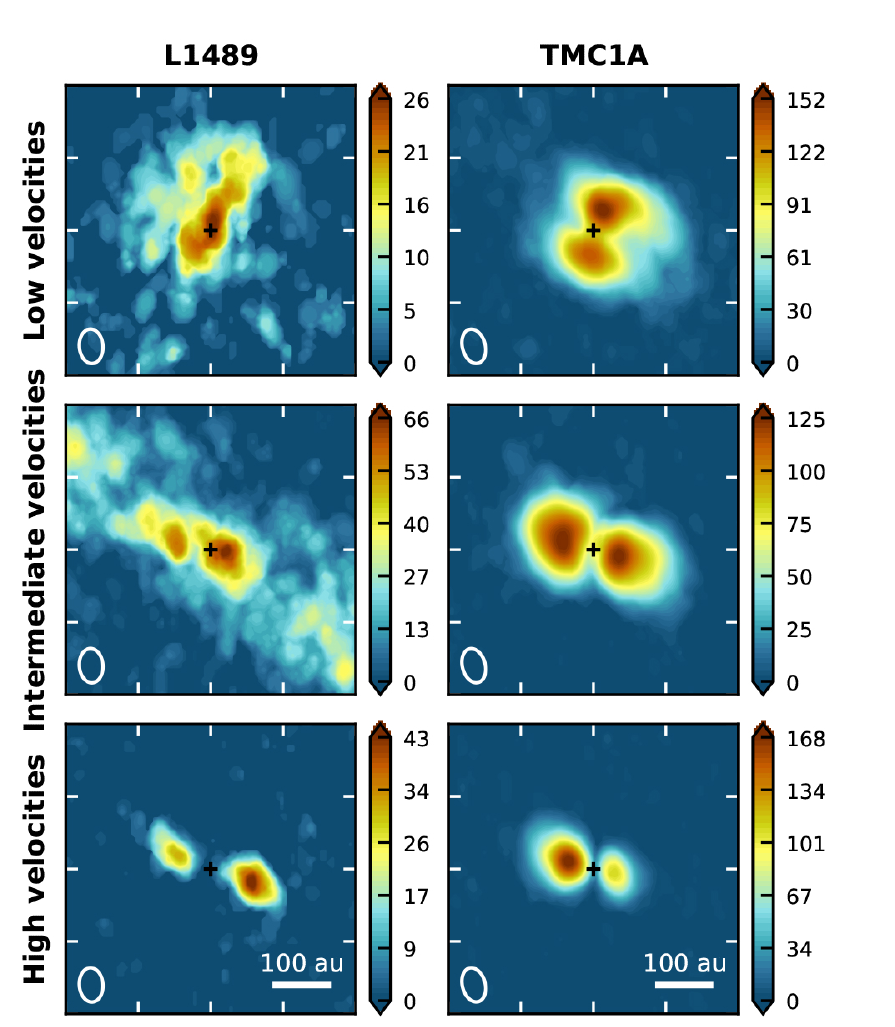}
\caption{Integrated intensity (moment zero) maps of the low-velocity (\textit{top row}), intermediate-velocity (\textit{middle row}), and high-velocity (\textit{bottom row}) C$^{17}$O $J=2-1$ emission toward L1489 (\textit{left column}) and TMC1A (\textit{right column}). Only pixels with $>3\sigma$ emission are included. For TMC1A (L1489), low velocities range from -1.27 to 1.26 (-0.47 to 0.43) km~s$^{-1}$, the intermediate velocities include $|\Delta v|$ = 1.34--2.49 (0.50--3.00) km~s$^{-1}$, and  the high velocities are $|\Delta v|$ = 2.57--4.94 (3.05--4.65) km~s$^{-1}$ with respect to the source velocity. The color scale is in mJy beam$^{-1}$ km s$^{-1}$ The source position is marked with a black cross and the beam is shown in the lower left corner of the panels. A 100 au scalebar is present in the bottom panels.}
\label{fig:M0_incldisks}
\end{figure}

A clear V-shaped pattern is also visible in the H$_2$CO integrated emission map for IRAS 04302 (Fig.~\ref{fig:M0_overview}). The V-shape starts at around 55 au ($\sim$1 beam offset from the continuum peak). If the reduction of H$_2$CO in the midplane is fully due to freeze-out, the snowline is then located around (or inward of) $\sim$25 au. In L1527, H$_2$CO emission also appears to come from surface layers, except in the outer disk (see Figs.~\ref{fig:M0_overview} and \ref{fig:Verticalprofiles}). The cold models show that CO emission from the envelope becomes comparable in strength to emission from the disk if CO is frozen out in a large part of the disk. Given that the envelope contribution is much larger in L1527 than in IRAS~04302, the emission peaking in the outer disk midplane is likely originating in the envelope. Instead of a clear V-shape, the emission in the inner region forms two bright lanes along the continuum position. A similar pattern is seen in the individual channels. This suggests that the H$_2$CO snowline is unresolved at the current resolution and closer in than in IRAS 04302 ($\lesssim$ 25 au). 

A zeroth order estimate of the midplane temperature profile for IRAS 04302 can be made from these two snowline estimates using a radial power law, $T \propto R^{-q}$. For disks, often a power law exponent $q$ of 0.5 is assumed, but $q$ can range between 0.33 and 0.75 \citep[see e.g.,][]{Adams1986,Kenyon1993a,Chiang1997}. A power law with $q$ = 0.75 matches the two temperature estimates reasonably well (see Fig.~\ref{fig:PowerlawT}). This temperature profile is quite similar to the profile constructed for L1527 based on $^{13}$CO and C$^{18}$O temperature measurements \citep{vantHoff2018b}. The L1527 temperature profile predicts a H$_2$CO snowline radius of $\lesssim$ 10 au, consistent with the results derived above. IRAS 04302 is thus warm like L1527, with freeze-out occuring only in the outermost part of this large disk. 

\begin{figure*}
\centering
\includegraphics[width=\textwidth,trim={0cm 0.2cm 0cm 0cm},clip]{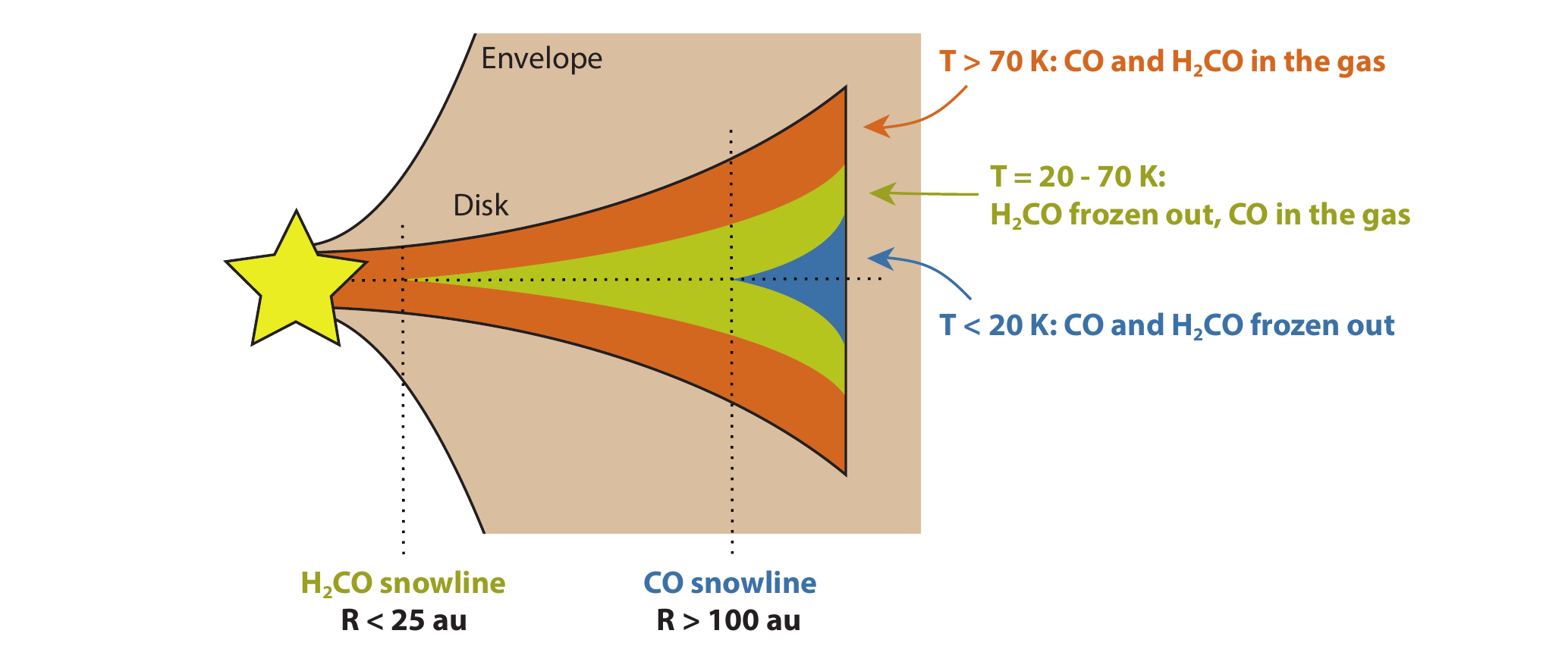}
\caption{Schematic representation of the temperature structure derived for Class I disks based on C$^{17}$O and H$_2$CO observations. A large part of the disk midplane, or even the entire midplane, is too warm for CO to freeze out unlike protoplanetary disks that have the CO snowline at a few tens of au. The majority of the midplane has a temperature between $\sim$20--70 K such that CO is in the gas phase while H$_2$CO is frozen out. The C$^{17}$O emission therefore arises predominantly from the midplane region (yellow area), and the H$_2$CO emission from the surface layers (orange region). }
\label{fig:CartoonTemperatureStructure}
\end{figure*}

\subsection{Temperature structure in less-inclined disks}

For less inclined disks, observing freeze-out directly is much harder; the projected area between the top and bottom layer becomes smaller (that is, the V-shape becomes more narrow), therefore requiring higher spatial resolution to observe it. In addition, because now both the near and the far side of the disk become visible, emission from the far side's surface layers can appear to come from the near side's midplane (see Figure~\ref{fig:incldisk}, and \citealt{Pinte2018}), which makes a V-shape due to emission originating only in the surface layers harder to observe. For the L1527 disk model, the intermediate and warm model become quite similar for an inclination of 60$^\circ$ at this angular resolution, and only a cold disk shows a clear V-shaped pattern (Fig.~\ref{fig:M0_incl-effect}). 

Figure~\ref{fig:M0_incldisks} shows the C$^{17}$O moment zero maps for the intermediate inclined disks TMC1A and L1489. The disk size, stellar mass and stellar luminosity of TMC1A are comparable to L1527. At intermediate velocities there is no sign of a V-shaped pattern, so these observations do not suggest substantial freeze out of CO in TMC1A. In order to constrain the CO snowline a little better, models were run with snowline locations of 31, 42 and 56 au (that is, in between the cold and intermediate model). All three models show a V-shape, suggesting that the CO snowline is at radii $\gtrsim$70 au in TMC1A. This is consistent with the results from \citet{Aso2015}, who found a temperature of 38 K at 100 au from fitting a disk model to ALMA C$^{18}$O observations, and the results from Harsono et al. (submitted), who find a temperature of 20 K at 115 au. There is no sign of a V-shaped pattern in the H$_2$CO emission. 

For L1489, the intermediate velocities show a more complex pattern with CO peaking close to the source and at larger offsets ($\gtrsim2^{\prime\prime}$). A similar structure was seen in C$^{18}$O \citep{Sai2020}. This could be the result of non-thermal desorption of CO ice in the outer disk if the dust column is low enough for UV photons to penetrate \citep{Cleeves2016}, or due to a radial temperature inversion resulting from radial drift and dust settling \citep{Facchini2017}. Such a double CO snowline has been observed for the protoplanetary disk IM Lup \citep{Oberg2015,Cleeves2016}. The structure of the continuum emission, a bright central part and a fainter outer part, make these plausible ideas. Another possibility is that the extended emission is due to a warm inner envelope component. The UV irradiated mass of L1489 derived from $^{13}$CO 6--5 emission is similar to that of L1527 and higher than for TMC1A and TMC1 \citep{Yildiz2015}. This may provide a sufficient column along the outflow cavity wall for C$^{17}$O emission to be observed. A high level of UV radiation is supported by O and H$_2$O line fluxes \citep{Karska2018}. 

If the edge of the compact CO emission is due to freeze-out, the CO snowline is located at roughly 200 au. Models based on the continuum emission have temperatures of $\sim$30 K or $\sim$20--30 K at 200 au (\citealt{Brinch2007,Sai2020}, respectively), so CO could indeed be frozen out in this region. The H$_2$CO emission does not show a gap at 200 au, which could mean that the emission is coming from the surface layers. The C$^{17}$O (and C$^{18}$O) abundance in these warmer surface layers may then be too low to be detected at the sensitivity of these observations.


\section{Discussion} \label{sec:Discussion}

\subsection{Temperature structure of young disks} 

We have used observations of C$^{17}$O and H$_2$CO toward five Class I disks in Taurus to address whether embedded disks are warmer than more evolved Class II disks. While the C$^{17}$O observations can indicate the presence or absence of $\lesssim$20 K gas, the addition of H$_2$CO observations allows to further constrain the temperature profile. The picture that is emerging suggests that these young disks have midplanes with temperatures between $\sim$20 and $\sim$70 K; cold enough for H$_2$CO to freeze out, but warm enough to retain CO in the gas phase (Fig.~\ref{fig:CartoonTemperatureStructure}). This suggests that, for example, the elemental C/O ratio in both the gas and ice could be different from that in protoplanetary disks. If planet formation starts during the embedded phase, the conditions for the first steps of grain growth are then thus different than generally assumed. 

Young disks being warmer than protoplanetary disks can also have consequences for the derived disk masses from continuum fluxes. This has been taken into consideration in recent literature by adopting a dust temperature of 30 K for solar-luminosity protostars \citep{Tobin2015a,Tobin2016b,Tychoniec2018,Tychoniec2020}, although not uniformly \citep[e.g.,][]{Williams2019,Andersen2019}, while 20 K is generally assumed for protoplanetary disks \citep[e.g.,][]{Ansdell2016}. In their study of Orion protostars, \citet{Tobin2020} take this one step further by scaling the temperature by luminosity based on a grid of radiative transfer models resulting in an average temperature of 43 K for a 1 $L_{\odot}$ protostar. Since higher temperatures will result in lower masses for a certain continuum flux, detailed knowledge of the average disk temperature is crucial to determine the mass reservoir available for planet formation. While the current study shows that embedded disks are warmer than protoplanetary disks, and the radial temperature profiles for L1527 and IRAS 04302 hint that 30 K may be to low for the average disk temperature, source specific modeling of the continuum and molecular line emission is required to address what would be an appropriate temperature to adopt for the mass derivation. However, an increase in temperature by a factor two will lower the mass by only a factor ~two (see Eq.~\ref{eq2}), and \citet{Tobin2020} still find embedded disks to be more massive than protoplanetary disks by a factor $>4$. Differences in temperature can thus not account for the mass difference observed between embedded and protoplanetary disks.

\subsubsection{The textbook example of IRAS 04302} 

The C$^{17}$O and H$_2$CO emission toward IRAS 04302 present a textbook example of what you would expect to observe for an edge-on disk, that is, a direct view of the vertical structure. The C$^{17}$O emission is confined to the midplane, while H$_2$CO is tracing the surface layers. Assuming the absence of H$_2$CO emission in the midplane is due to freeze out, we can not only make a first estimate of the radial temperature profile but also of the vertical temperature structure. At the current spatial resolution, the vertical structure is spatially resolved for radii $\gtrsim70$ au, that is, $\sim$3 beams across the disk height. At these radii, the H$_2$CO emission peaks $\sim$30--50 au above the midplane (at radii of 70 and 180 au, respectively), suggesting that the temperature is between $\sim$20--70 K in the $\sim$30 au above the midplane.  

The temperature structure can be further constrained by observing molecules with a freeze-out temperature between that of CO and H$_2$CO, that is, between $\sim$20--70 K. Based on the UMIST database for astrochemistry \citep{McElroy2013}, examples of such molecules are CN, CS, HCN, C$_2$H, SO and H$_2$CS (in increasing order of freeze-out temperature). Another option would be to observe several H$_2$CO lines because their line ratios are a good indicator of the temperature \citep[e.g.,][]{Mangum1993}. These observations thus confirm that edge-on disks are well-suited to study the disk vertical structure through molecular line observations. 

\subsubsection{Comparison with protostellar envelopes and protoplanetary disks}

\begin{figure}
\centering
\includegraphics[width=0.5\textwidth,trim={0cm 16.3cm 7.5cm .8cm},clip]{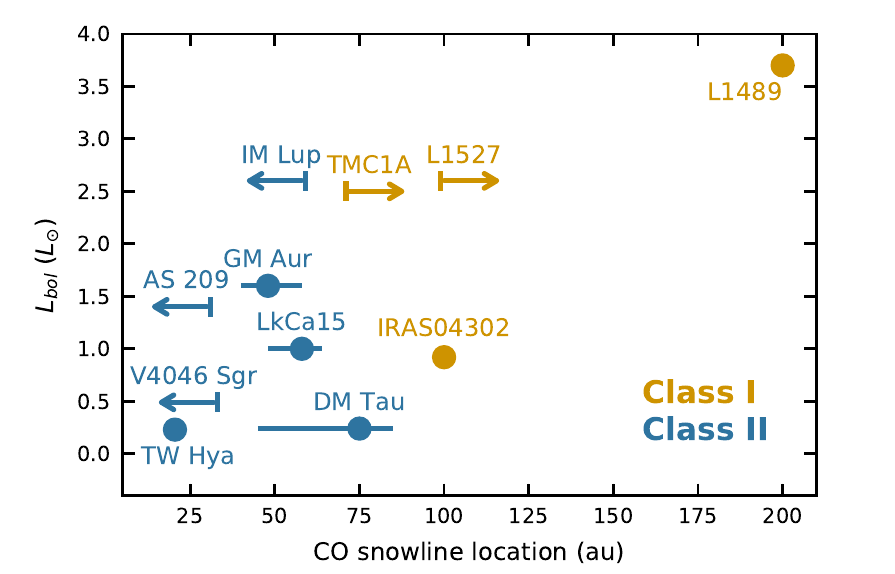}
\caption{Overview of CO snowline locations in disks derived from molecular line observations as function of bolometric luminosity. The locations for Class I disks (orange) are derived in this work using the C$^{17}$O emission. Class II T Tauri disks are shown in blue. For TW Hya, the CO snowline location is determined from $^{13}$C$^{18}$O emission by \citet{Zhang2017}. For the other Class II disks, the CO snowline is derived from N$_2$H$^+$ emission by \citet{Qi2019}. Arrows denote upper and lower limits.  }
\label{fig:COsnowlineLocations}
\end{figure}

No sign of CO freeze-out is detected in the C$^{17}$O observations of L1527, and while freeze-out is much more difficult to see in non-edge on disks, TMC1A does not show hints of freeze out at radii smaller than $\sim$70 au. A first estimate puts the CO snowline at $\sim$100 au in IRAS~04302, and the CO snowline may be located around $\sim$200 AU in L1489. These young disks are thus warmer than T Tauri disks where the snowline is typically at a few tens of AU, as can be seen in Fig.~\ref{fig:COsnowlineLocations}. We only include Class II disks for which a CO snowline location has been reported based on molecular line observations, either $^{13}$C$^{18}$O (for TW Hya; \citealt{Zhang2017}) or N$_2$H$^+$ \citep{Qi2019}. There is no clear trend between CO snowline location and bolometric luminosity for either Class, but the Class I disks have CO snowlines at larger radii compared to Class II disks with similar bolometric luminosities. 

In protostellar envelopes, snowline radii larger than expected based on the luminosity have been interpreted as a sign of a recent accretion burst \citep{Jorgensen2015,Frimann2017,Hsieh2019}. During such time period of increased accretion, the circumstellar material heats up, shifting the snowlines outward. Once the protostar returns to its quiescent stage, the temperature adopts almost instantaneously, while the chemistry takes longer to react. During this phase the snowlines are at larger radii than expected from the luminosity. The results in Fig.~\ref{fig:COsnowlineLocations} could thus indicate that small accretion bursts have occurred in the Class I systems and that the CO snowlines have not yet shifted back to their quiescent location. When such a burst should have happened depends on the freeze-out timescale, $\tau_{\rm{fr}}$; 
\begin{equation}
\tau_{\rm{fr}} = 1 \times 10^4 \hspace{1mm} \rm{yr} \sqrt{\frac{10 \hspace{1mm} \rm{K}}{T_{\rm{fr}}}} \frac{10^6 \hspace{1mm} \rm{cm}^{-3}}{n_{\rm{H_2}}},
\end{equation}
where $T_{\rm{fr}}$ is the freeze-out temperature and $n_{\rm{H_2}}$ is the gas density \citep{Visser2012}. For densities of $\gtrsim 10^8$ cm$^{-3}$, the CO freeze out timescale is $\lesssim$ 100 yr. This could suggest that Class I protostars frequently undergo small accretion bursts. Alternatively, these young disks may have lower densities than more evolved disks. As shown by the model results from Murillo et al. (in preperation), decreasing the density while keeping the luminosity constant shifts the snowlines outward. If this is what is causing the results in Fig.~\ref{fig:COsnowlineLocations}, this means that embedded disks not only have different temperature structures from protoplanetary disks, but also different density structures. However, the larger disk masses derived for embedded disks compared to protoplanetary disks for similar disk radii makes this unlikely \citep{Tobin2020}.

Another comparison is made in Fig.~\ref{fig:PowerlawT}, where the radial temperature profiles inferred for L1527 and IRAS~04302 are shown together with those for the younger Class 0 disk-like structure around IRAS~16293A \citep{vantHoff2020} and the Class II disk TW Hya \citep{Schwarz2016}. The young disks are warmer than the more evolved Class~II disk, but much colder than the Class~0 system IRAS~16293A. When making this comparison one should keep in mind that IRAS~16293A reflects an envelope where the temperature will be larger at larger scales because of the spherical rather than disk structure. In a disk the temperature will drop more rapidly in the radial direction due to the higher extinction compared to an envelope. Nevertheless, such an evolutionary trend is expected because the accretion rate decreases as the envelope and disk dissipate. As a consequence, heating due to viscous accretion diminishes and hence the temperature drops, as shown by two-dimensional physical and radiative transfer models for embedded protostars \citep{D'Alessio1997,Harsono2015}. In addition, the blanketing effect of the envelope decreases as the envelope dissipates \citep{Whitney2003}. 

As a first comparison between the observations and model predictions, models from \citet{Harsono2015} are overlaid on the observationally inferred temperature profiles in Fig.~\ref{fig:PowerlawT} (right panel). In these models the dust temperature is determined based on stellar irradiation and viscous accretion. Models are shown for a stellar luminosity of 1 $L_\odot$, an envelope mass of 1 $M_\odot$, a disk mass of 0.05 $M_\odot$, a disk radius of 200 au and different accretion rates. The disk mass has a negligible effect on the temperature profiles (see \citealt{Harsono2015} for details). IRAS 16293A matches reasonably well with the temperature profile for a heavily accreting system ($10^{-4} M_\odot$ yr$^{-1}$), consistent with estimates of the accretion rate (e.g., \mbox{$\sim 5 \times 10^{-5} M_\odot$ yr$^{-1}$;} \citealt{Schoier2002}). However, as in these models the total luminosity is based on the stellar luminosity and the accretion luminosity (and a contribution from the disk), the match for IRAS 16239A with a strongly accretion model may just reflect the systems bolometric luminosity of 20 $L_\odot$. In contrast, the temperature profiles for L1527 and IRAS 04302 are comparable to the colder \mbox{$10^{-7} M_\odot$ yr$^{-1}$} model, consistent with the accretion rate of $\sim 3 \times 10^{-7} M_\odot$ yr$^{-1}$ for L1527 (see \citealt{vantHoff2018b}). Similar accretion rates in the order of $10^{-7} M_\odot$ yr$^{-1}$ have been reported for L1489, TMC1A and TMC1 \citep[e.g.,][]{Mottram2017,Yen2017} based on the bolometric luminosities \citep[see e.g.,][]{Stahler1980,Palla1993}. We are not aware of a measurement toward IRAS 04302, but our very preliminary modeling results (van 't Hoff et al. in prep.) are consistent with an accretion rate in the order of $10^{-7} M_\odot$ yr$^{-1}$. Measured accretion rates for TW Hya range between $\sim$ $2 \times 10^{-10} - 2 \times 10^{-9} M_\odot$ yr$^{-1}$ \citep[e.g.,][]{Herczeg2008,Curran2011,Ingleby2013}, and accretion rates of $\sim$ $10^{-10} - 10^{-8} M_\odot$ yr$^{-1}$ are typically measured for protoplanetary disks around T Tauri stars (see \citealt{Hartmann2016} for a review).

The results presented here thus provide observational evidence for cooling of the circumstellar material during evolution. More sources need to be observed to confirm this trend and to answer more detailed questions such as, when has a disk cooled down sufficiently for large-scale CO freeze-out? Does this already happen before the envelope dissipates? IRAS 04302 is a borderline Class I/Class II object embedded in the last remnants of its envelope, but still has a temperature profile more similar to L1527 than TW Hya. Although a caveat here may be the old age of TW Hya ($\sim$10 Myr), this hints that disks may stay warm until the envelope has fully dissipated. 

\subsubsection{TMC1} 

TMC1 is for the first time resolved to be a close ($\sim$85~au) binary. A possible configuration of the system could be that TMC1-E is present in the disk of TMC1-W, as for example observed for L1448 IRS3B \citep{Tobin2016a}. TMC1-E would then increase the temperature in the east side of the disk. This may be an explanation for the asymmetry in the C$^{17}$O emission with the emission dimmer east of TMC1-W (see Figs.~\ref{fig:Radprofiles}, \ref{fig:M1_overview} and \ref{fig:Spectra}). Given the upper level energy of 16 K, emission from the C$^{17}$O $J=2-1$ transition will decrease with temperatures increasing above $\sim$25~K. The weaker C$^{17}$O emission may thus signal a higher temperature in the east side of the disk. However, TMC1-E does not seem to cause any disturbances in the disk, such as spiral arms, although the high inclination may make this hard to see. Another possibility could be that TMC1-E is actually in front of the disk.   

\subsection{Chemical complexity in young disks} 

One of the major questions regarding the chemical composition of planetary material, is whether they contain complex organic molecules (COMs). Due to the low temperatures in protoplanetary disks, observations of COMs are very challenging because these molecules thermally desorb at temperatures $\gtrsim$100--150 K, that is, in the inner few AU. In contrast, COMs are readily detected on disk-scales in protostellar envelopes (e.g., IRAS 16293, NGC1333 IRAS2A, NGC1333 IRAS4A, and B1-c; \citealt{Jorgensen2016,Taquet2015,vanGelder2020}) and in the young disk V883-Ori, where a luminosity outburst has heated the disk and liberated the COMs from the ice mantles \citep{vantHoff2018c,Lee2019}. 

Although young disks seem warmer than protoplanetary disks, the CH$_3$OH and HDO non-detections with upper limits orders of magnitude below column densities observed toward Class 0 protostellar envelopes suggest that they are not warm enough to have a hot-core like region with a large gas reservoir of COMs. This is consistent with recent findings by \citet{ArturdelaVillarmois2019} for a sample of Class I protostars in Ophiuchus. More stringent upper limits are required for comparison with the Class II disks TW Hya and HD 163296. However, the detection of HDO and CH$_3$OH may have been hindered by optically thick dust in the inner region, or by the high inclinations of these sources. Modeling by Murillo et al. (in preperation) shows that the water snowline is very hard to detect in near edge-on disks. These non-detections thus do not rule out the presence of HDO and CH$_3$OH, in fact, if the region where HDO and CH$_3$OH are present is much smaller than the beam, they may have higher columns than the upper limits derived here. This is corroborated by the weak detection of CH$_3$OH in L1527 \citep{Sakai2014b}. These results thus merely show that Class I disks do not have an extended hot-core like region, making the detection of COMs just as challenging as in Class II disks.

A related question to the chemical composition is whether the disk material is directly inherited from the cloud, processed en route to the disk, or even fully reset upon entering the disk. Young disks like L1527, where no CO freeze-out is observed, suggest that no full inheritance takes place, at least not for the most volatile species like CO. Ice in the outer disk of IRAS 04302 could be inherited. However, the freeze-out timescale for densities $> 10^6$ cm$^{-3}$ is $< 10^4$ year, so this CO could have sublimated upon entering the disk and frozen out as the disk cooled (see e.g., \citealt{Visser2009a}). Without CO ice, additional grain-surface formation of COMs will be limited in the young disks. So if COMs are present in more evolved disks, as for example shown for V883 Ori, they must have been inherited from a colder pre-collapse phase. Physicochemical models show that prestellar methanol can indeed be incorporated into the disk \citep{Drozdovskaya2014}.

\subsection{Decrease in H$_2$CO in the inner disk} 

While the H$_2$CO emission is brighter than the C$^{17}$O emission at intermediate velocities, no H$_2$CO emission is detected at the highest velocities in IRAS 04302, L1527 and TMC1A, suggesting a reduction in H$_2$CO flux in the inner $\lesssim$20--30 au in these disks. This is not just a sensitivity issue, as for example, C$^{17}$O and H$_2$CO have similar strength and emitting area in channels around +1.9 km s$^{-1}$ with respect to the source velocity in L1527 while 3.05 km s$^{-1}$ is the highest velocity observed for C$^{17}$O and 2.60 km s$^{-1}$ the highest velocity for H$_2$CO. The decrease in H$_2$CO emission is also unlikely to be due to the continuum being optically thick because this would affect the C$^{17}$O emission as well, unless there is significantly more C$^{17}$O emission coming from layers above the dust millimeter $\tau$ = 1 surface than H$_2$CO emission. Given the observed distributions with H$_2$CO being vertically more extended than C$^{17}$O this seems not to be the case. Moreover, the drop in H$_2$CO in TMC1A occurs much further out than where the dust becomes optically thick. 

Formaldehyde rings have also been observed in the protoplanetary disks around TW Hya \citep{Oberg2017}, HD 163296 \citep{Qi2013b,Carney2017}, DM Tau \citep{Henning2008,Loomis2015} and DG Tau \citep{Podio2019}. Interestingly, a ring is only observed for the $3_{03}-2_{02}$ and $3_{12}-2_{11}$ transitions and not for the $5_{15}-4_{14}$ transition. \citet{Oberg2017} argue that the dust opacity cannot be the major contributor in TW Hya, because the dust opacity should be higher at higher frequencies, thus for the $5_{15}-4_{14}$ transition. Instead, they suggest a warm inner component that is visible in the $5_{15}-4_{14}$ transition ($E_{\mathrm{up}}$ = 63 K) and not in the $3_{12}-2_{11}$ transition ($E_{\mathrm{up}}$ = 33 K). For L1527, we observe the $3_{12}-2_{11}$ transition and radiative transfer modeling for the L1527 warm disk model shows that both the C$^{17}$O ($E_{\mathrm{up}}$ = 33 K) and H$_2$CO emission goes down by a factor $\sim$2 if the temperature is increased by 80\%. An excitation effect thus seems unlikely, unless the C$^{17}$O emission is optically thick. The latter is not expected given that the C$^{18}$O in L1527 is only marginally optically thick \citep{vantHoff2018b}. The absence of H$_2$CO emission in the inner disk thus points to a reduced H$_2$CO abundance. A lower total (gas + ice) H$_2$CO abundance (more than an order of magnitude) in the inner 30 au is seen in models by \citet{Visser2011}, who studied the chemical evolution from pre-stellar core into disk, but these authors do not discuss the H$_2$CO chemistry. 

The H$_2$CO abundance in the inner disk can be low if its formation is inefficient. H$_2$CO can form both in the gas and in the ice \citep[e.g.,][]{Willacy2009,Walsh2014,Loomis2015}. On the grain surfaces, the dominant formation route is through hydrogenation of CO \citep{Watanabe2002,Cuppen2009,Fuchs2009}. Since there seems to be no CO freeze out in these young disks, or only at radii $\gtrsim$ 100 au, H$_2$CO is expected to form predominantly in the gas. Ring-shaped H$_2$CO emission due to increased ice formation outside the CO snowline, as used to explain the ring observed in HD 163296 \citep{Qi2013b}, is thus not applicable to the disks in this sample. 

In the gas, the reaction between CH$_3$ and O is the most efficient way to form H$_2$CO \citep[e.g.,][]{Loomis2015}. Therefore, a decrease in gas-phase H$_2$CO formation would require a low abundance of either CH$_3$ or O. CH$_3$ is efficiently produced by photodissociation of CH$_4$ or through ion-molecule reactions. A low CH$_3$ abundance thus necessitates the majority of carbon to be present in CO, in combination with a low X-ray flux as carbon can only be liberated from CO by X-ray generated He$^+$. Atomic oxygen is formed through photodissociation H$_2$O and CO$_2$, or through dissociation of CO via X-ray-generated He$^+$. A low atomic oxygen abundance would thus require a low UV and X-ray flux.

Besides a low formation rate, a high destruction rate would also decrease the amount of H$_2$CO. However, the destruction products have a limited chemistry and re-creation of H$_2$CO is the most likely outcome. \citet{Willacy2009} showed that a third of the ions formed by H$_2$CO destruction through HCO$^+$ and DCO$^+$ form CO instead of reforming H$_2$CO, leading to a depletion between 7 and 20 AU for their disk model. However, this only reduces H$_2$CO in the midplane, not in the surface layers. In addition, \citet{Henning2008} suggested the conversion of CO into CO$_2$-containing molecules and hydrocarbons that freeze out onto dust grains (see also \citealt{Aikawa1999}). However, the C$^{17}$O observations do not suggest heavy CO depletion. 

Another effect that could contribute is photodesorption of methanol ice that is inherited from earlier phases. Laboratory experiments have shown that methanol does not desorb intact upon VUV (vacuum ultraviolet) irradiation, but rather leads to the release of smaller photofragments including H$_2$CO \citep{Bertin2016,Cruz-Diaz2016}. This could lead to an increase of H$_2$CO outside the region where CH$_3$OH ice thermally desorbs ($\sim100-150$ K). Finally, turbulence may play a role as models by \citet{Furuya2014} show the formation of H$_2$CO rings when mixing is included. However, these rings are due to a decrease of H$_2$CO inside the CO snowline and an increase outside this snowline, and these results may not be applicable to embedded disks without CO freeze out. Observations of higher excitation H$_2$CO lines and chemical modeling with source-specific structures may provide further insights. 

It is worth noting that \citet{Pegues2020} found both centrally-peaked and centrally-depressed H$_2$CO emission profiles for a sample of 15 protoplanetary disks. A reduction of H$_2$CO emission toward three out of the five disks in our sample could mean that the H$_2$CO distribution is set during the embedded stage.


\section{Conclusions}  \label{sec:Conclusions}

Temperature plays a key role in the physical and chemical evolution of circumstellar disks, and therefore in the outcome of planet formation. However, the temperature structure of young embedded disks, in which the first steps of planet formation take place, is poorly constrained. Our previous analyis of $^{13}$CO and C$^{18}$O emission in the young disk L1527 suggest that this disk is warm enough ($T \gtrsim$ 20-25 K) to prevent CO freeze-out \citep{vantHoff2018b} in contrast to protoplanetary disks that show large cold outer regions where CO is frozen out. Here we present ALMA observations of C$^{17}$O, H$_2$CO and non-detections of HDO and CH$_3$OH for five young disks in Taurus, including L1527. The observations of L1527 and in particular IRAS~04302, with C$^{17}$O emission originating in the midplane and H$_2$CO emission tracing the surface layers, highlight the potential of edge-on disks to study the disk vertical structure. 

Based on the following results we conclude that young disks are likely warmer than more evolved protoplanetary disks, but not warm enough to have a large gas reservoir of complex molecules, like the young disk around the outbursting star V883-Ori:

\begin{itemize}
\item CO freeze-out can be observed directly with C$^{17}$O observations in edge-on disks. L1527 shows no sign of CO freeze-out, but IRAS 04302 has a large enough disk for the temperature to drop below the CO freeze-out temperature in the outermost part (radii $\gtrsim$100 au). 

\item H$_2$CO emission originates primarily in the surface layers of IRAS~04302 and L1527. The snowline ($T \sim$70 K) is estimated around (or inward of) $\sim$25 au in IRAS~04302 and at $\lesssim$25 au in L1527.

\item CO freeze-out is much more difficult to observe in non-edge-on disks, but the C$^{17}$O emission in TMC1A suggest a snowline at radii $\gtrsim$ 70 au. Two spatial components are seen in the C$^{17}$O emission toward L1489. If the outer edge of the inner component is due to CO freeze-out, the snowline would be around $\sim$200 au. 

\item The CO snowline locations derived for the Class I disks are farther out than found for Class II disks with similar bolometric luminosities. 

\item The HDO and CH$_3$OH non-detections with upper limits more than two orders of magnitude lower than observed for hot cores in protostellar envelopes or the disk around the outbursting star V883-Ori suggest that these Class I disks do not have a large gas reservoir of COMs. 

\item The inferred temperature profiles are consistent with trends found in radiative transfer models of disk-envelope systems with accretion rates decreasing from $10^{-4}$ to $10^{-7} M_\odot$ yr$^{-1}$. 
\end{itemize} 

\noindent As evidence is piling up for planet formation to start already during the embedded phase, adopting initial conditions based on the physical conditions in more evolved Class II disks seems not appropriate. Instead, planet formation may start in warmer conditions than generally assumed. Furthermore, without a large CO-ice reservoir, COM formation efficiency is limited in embedded disks. Observations of COMs in more evolved disks therefore suggest that these molecules are inherited from earlier phases.


\acknowledgements

We would like to thank the referee for a prompt and positive report that helped improve the paper, Patrick Sheehan for his assistence with the visibility plotting and Gleb Fedoseev for useful discussions about the H$_2$CO freeze-out temperature. M.L.R.H. would like to thank Yuri Aikawa for comments on an earlier version of this manuscript for her PhD thesis. This paper makes use of the following ALMA data: ADS/JAO.ALMA\#2017.1.01413.S. ALMA is a partnership of ESO (representing its member states), NSF (USA) and NINS (Japan), together with NRC (Canada), MOST and ASIAA (Taiwan), and KASI (Republic of Korea), in cooperation with the Republic of Chile. The Joint ALMA Observatory is operated by ESO, AUI/NRAO and NAOJ. Astrochemistry in Leiden is supported by the Netherlands Research School for Astronomy (NOVA). M.L.R.H acknowledges support from a Huygens fellowship from Leiden University. J.J.T. acknowledges support from grant AST-1814762 from the National Science Foundation and past support from the Homer L. Dodge Endowed Chair at the University of Oklahoma.The National Radio Astronomy Observatory is a facility of the National Science Foundation operated under cooperative agreement by Associated Universities, Inc. J.K.J acknowledges support by the European Research Council (ERC) under the European Union's Horizon 2020 research and innovation programme through ERC Consolidator Grant ``S4F'' (grant agreement No~646908). A.M. acknowledges funding from the European Union’s Horizon 2020 research and innovation programme under the Marie Sklodowska-Curie grant agreement No 823823, (RISE DUSTBUSTERS) and from the Deutsche Forschungsgemeinschaft (DFG, German Research Foundation) – Ref no. FOR 2634/1 ER685/11-1. C.W. acknowledges financial support from the University of Leeds and from the Science and Technology Facilities Council (grant numbers ST/R000549/1 and ST/T000287/1).


\bibliography{References}{}
\bibliographystyle{aasjournal}


\restartappendixnumbering

\begin{appendix}


\section{Observations} \label{ap:Observations}

\begin{figure*}
\centering
\includegraphics[width=\textwidth,trim={0cm 12.8cm 0cm 1cm},clip]{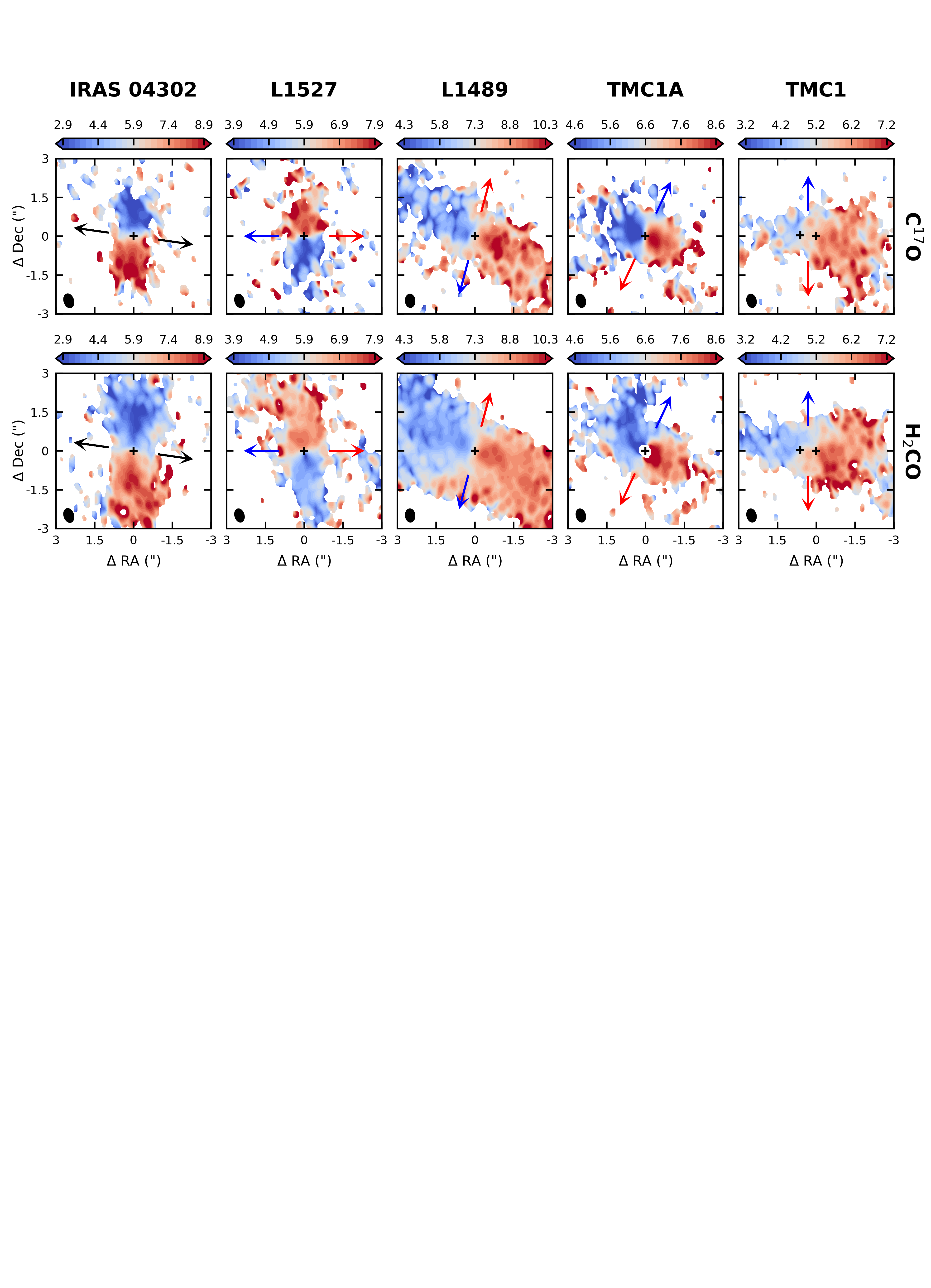}
\caption{Moment one maps for the C$^{17}$O $J=2-1$ (\textit{top row}) and H$_2$CO 3$_{1,2}$-2$_{1,1}$ (\textit{bottom row}) transitions. The central velocity of the color scale is the systemic velocity (km s$^{-1}$). The positions of the continuum peaks are marked with black crosses, and the outflow directions are indicated by arrows. The beam is shown in the lower left corner of each panel.}
\label{fig:M1_overview}
\end{figure*}

\begin{figure}
\centering
\includegraphics[trim={0cm 3cm 6.5cm .8cm},clip]{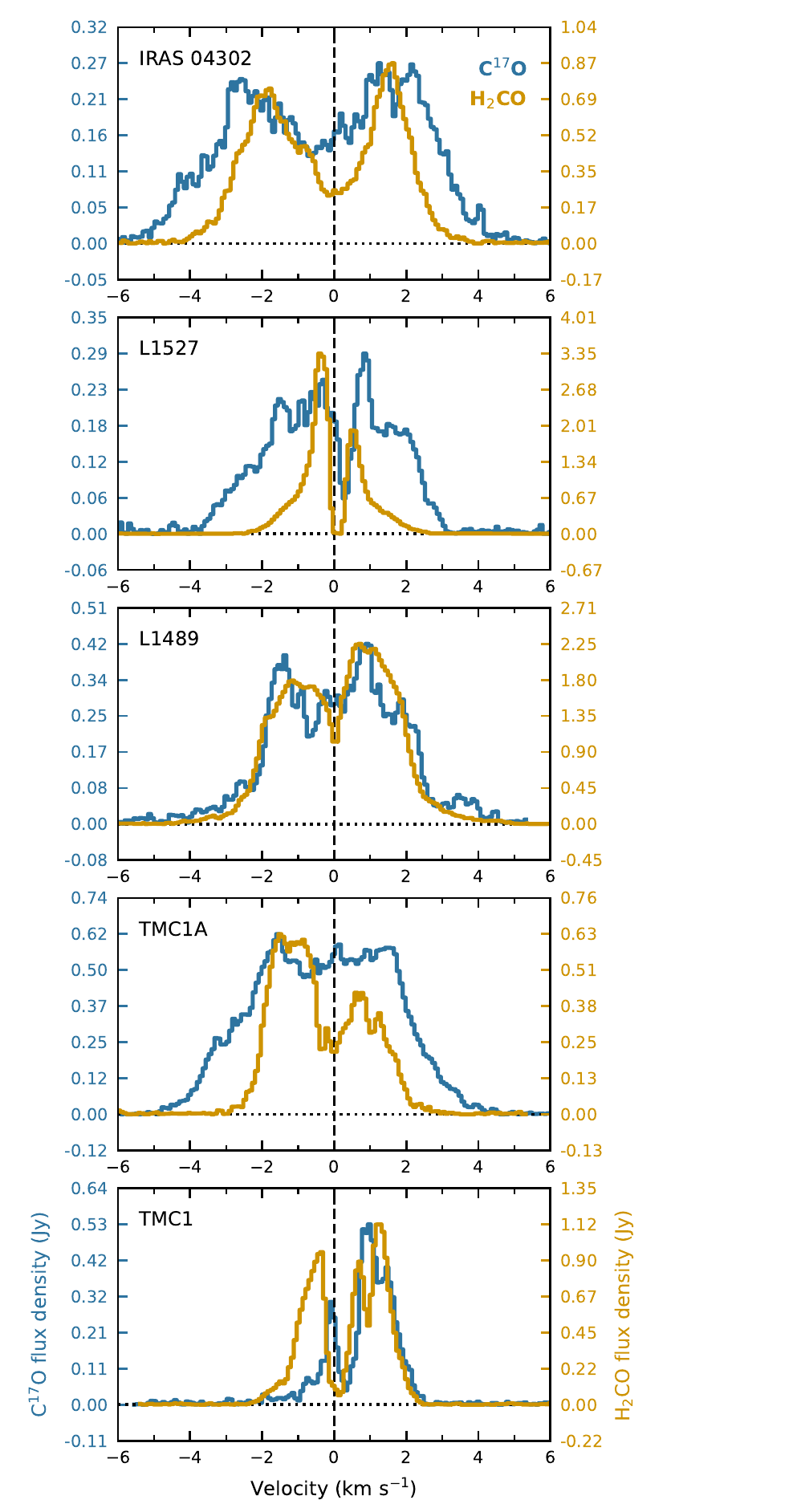}
\caption{Spectra for C$^{17}$O (blue) and H$_2$CO (orange) extracted in a 6$^{\prime\prime}$ circular aperture centered at the continuum peak. Only pixels with $>$ 3$\sigma$ emission are included. The vertical scale is different for each molecular line in each panel. The vertical dashed lines mark the systemic velocities, which have been shifted to 0 km s$^{-1}$.}
\label{fig:Spectra}
\end{figure}

Table~\ref{tab:Lineparameters} presents an overview of the observed molecular lines. Moment one maps for C$^{17}$O and H$_2$CO toward all disks in the sample are shown in Fig.~\ref{fig:M1_overview}, and spectra integrated over pixels with $>$ 3$\sigma$ emission in a 6$^{\prime\prime}$ circular aperture are presented in Fig.~\ref{fig:Spectra}.

\begin{table}
\caption{Overview of the molecular line observations.}
\label{tab:Lineparameters} 
\centering
\begin{footnotesize}
\begin{tabular}{l c c c c c c c c c}
    \hline\hline
    \\[-.3cm]
    Molecule & Transition & Frequency & $A_{\rm{ul}}$\tablenotemark{a} & $E_{\rm{up}}$\tablenotemark{b} &  \\ 
    & & (GHz) & (s$^{-1}$) & (K)   \\
    \hline 
    \\[-.3cm]
    C$^{17}$O & $2-1$ & 224.714385 & 6.42 $\times 10^{-7}$ & 16   \\
    H$_2$CO & $3_{1,2}-2_{1,1}$ & 225.697775 & 2.77 $\times 10^{-4}$ & 33   \\
    HDO & $3_{1,2}-2_{2,1}$  & 225.896720 & 1.32 $\times 10^{-5}$ & 168 \\
    CH$_3$OH & \hspace{2mm}$5-4$\tablenotemark{c} & 241.820762\tablenotemark{d} & 2--6 $\times 10^{-5}$ & 34--131\\
    \hline
\end{tabular}
\begin{flushleft}
\vspace{-0.3cm}
\tablecomments{Data for C$^{17}$O and HDO are taken from the Jet Propulsion Laboratory Molecular Spectroscopy database (JPL; \citealt{Pickett1998}), and data for H$_2$CO and CH$_3$OH are from the Cologne Database for Molecular Spectroscopy (CDMS; \citealt{Muller2005}).}
\vspace{-0.2cm} 
\tablenotetext{a}{Einstein A coefficient.}
\vspace{-0.2cm} 
\tablenotetext{b}{Upper level energy.}
\vspace{-0.2cm} 
\tablenotetext{c}{The spectral window covers multiple transitons in the $5_K-4_K$ branch for both A- and E-methanol (16 transitions in total).}
\vspace{-0.2cm} 
\tablenotetext{d}{Central frequency of the spectral window.}
\end{flushleft}
\end{footnotesize}
\end{table}


\section{Envelope contribution} \label{ap:Vismod}

\begin{figure*}
\centering
\includegraphics[width=\textwidth,trim={0cm 0cm 0cm 0cm},clip]{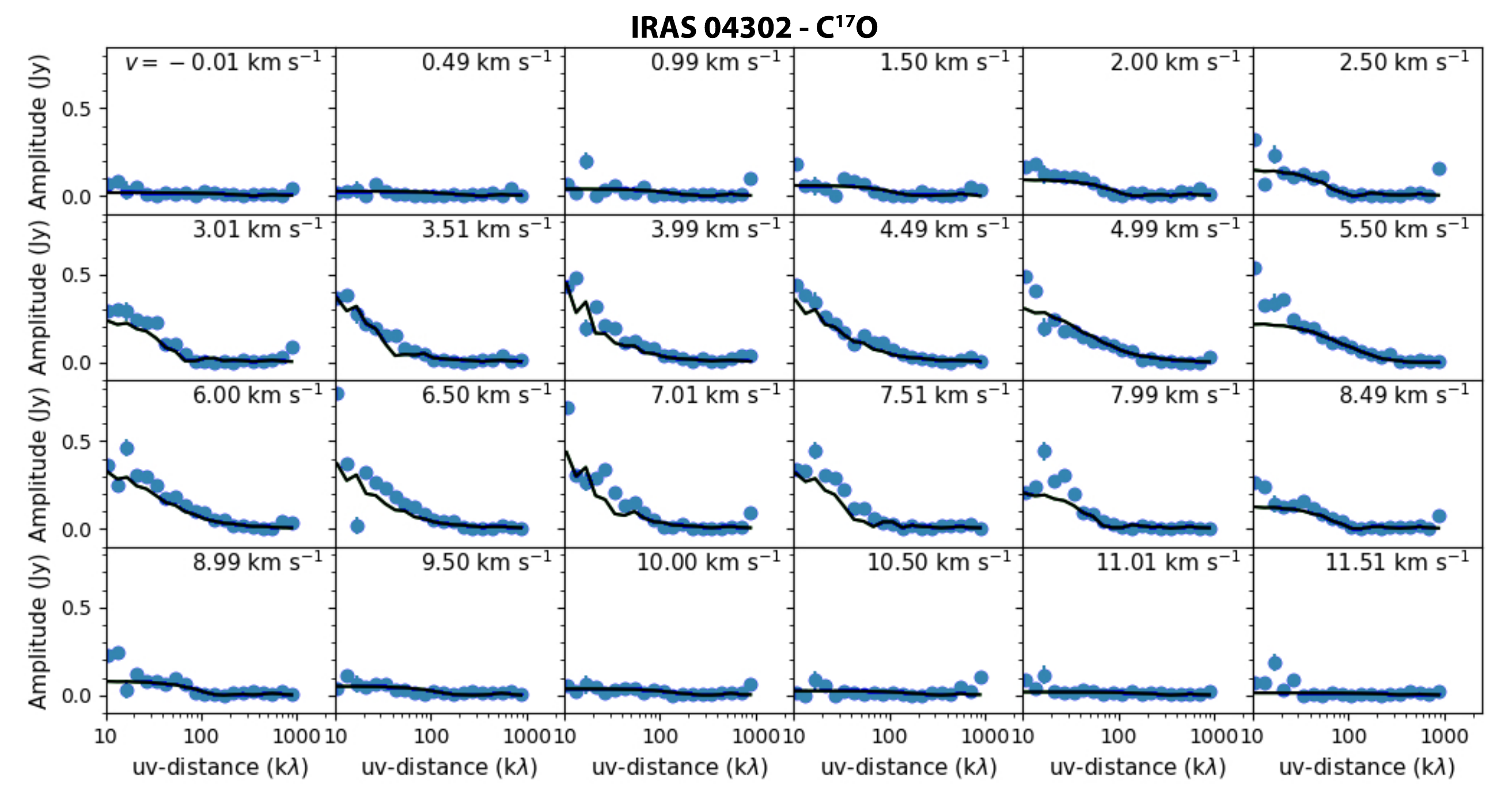}
\includegraphics[width=\textwidth,trim={0cm 0cm 0cm 0cm},clip]{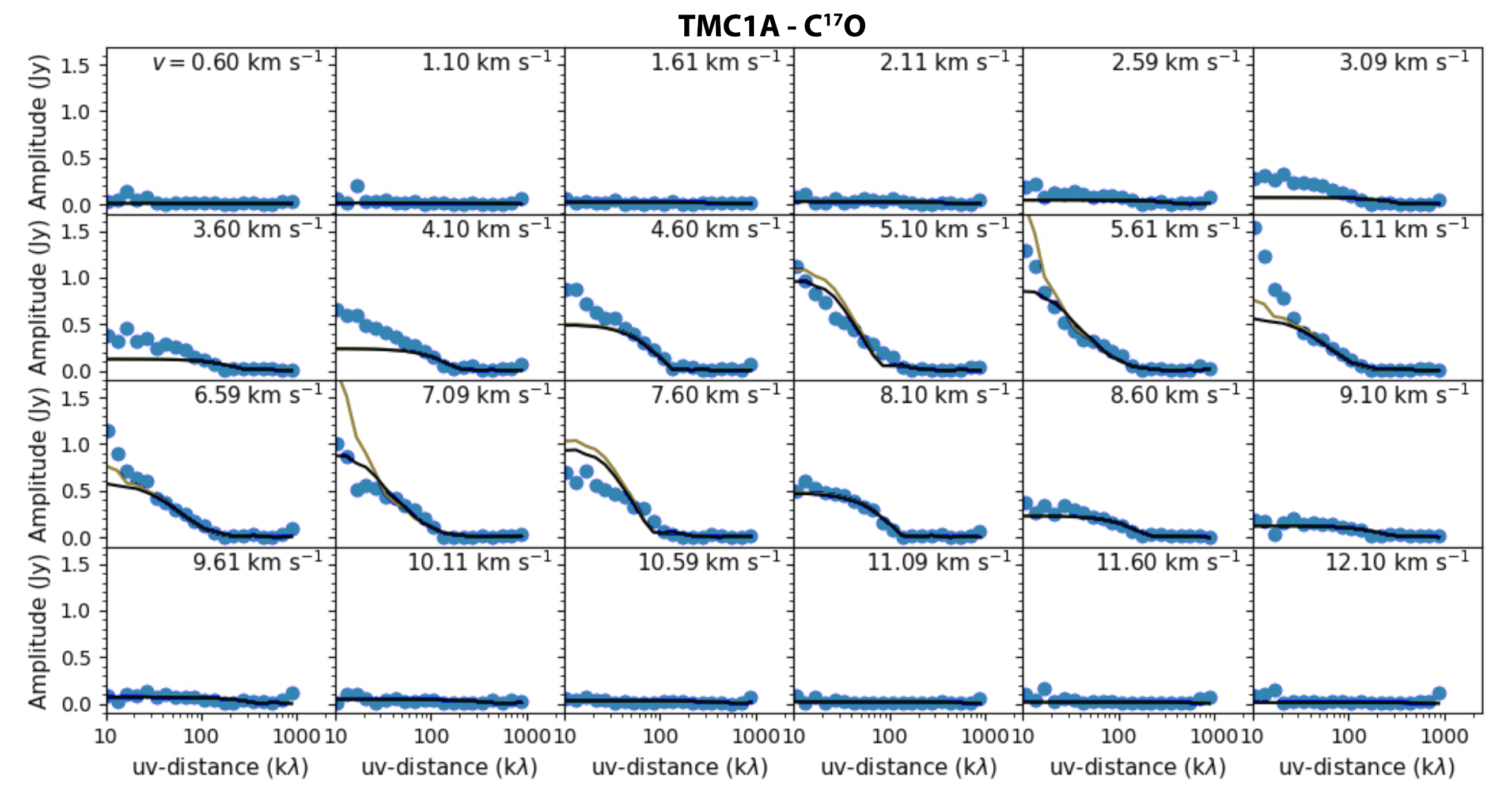}
\caption{Visibility amplitude profiles for C$^{17}$O toward IRAS 04302 (\textit{top panels}) and TMC1A (\textit{bottom panels}). The black line displays a Keplerian disk and the orange line represents a Keplerian disk plus rotating infalling envelope \citep{Ulrich1976}. The systemic velocities are 5.9 and 6.6 km s$^{-1}$ for IRAS 04302 and TMC1A, respectively.}
\label{fig:Visibilities}
\end{figure*}

A first assessment of the envelope contribution to the line emission can be made by comparing generic models of either only a Keplerian disk or a disk embedded in an envelope to the observed visibility amplitudes. To do so, we calculated the visibility amplitude profiles for a Keplerian disk in 0.5 km s$^{-1}$ channels using the modeling tools outlined in \citet{Sheehan2019}. Values for the stellar mass, disk radius, inclination and position angle were adopted from the literature, and the C$^{17}$O and H$_2$CO abundance were taken constant throughout the disk. The disk mass was adjusted to approximately match the visibility amplitude profiles in each channel. If there was a component at small uv-distances that could not be reproduced with the disk, we added a rotating infalling envelope with 3000 au radius using the prescription by \citet{Ulrich1976}. The results for C$^{17}$O toward IRAS 04302 and TMC1A are shown as an example in Fig.~\ref{fig:Visibilities}. We stress that we do not expect a perfect fit with this simple approach, but it shows that the C$^{17}$O emission toward IRAS 04302 can be reproduced without an envelope, while some envelope contribution is required at low velocities ($\sim |1|$ km s$^{-1}$ from the systemic velocity) toward TMC1A.


\section{Schematics of the disk models} \label{ap:Cartoons}

\begin{figure}
\centering
\includegraphics[width=0.5\textwidth,trim={0cm 2.5cm 0cm 2cm},clip]{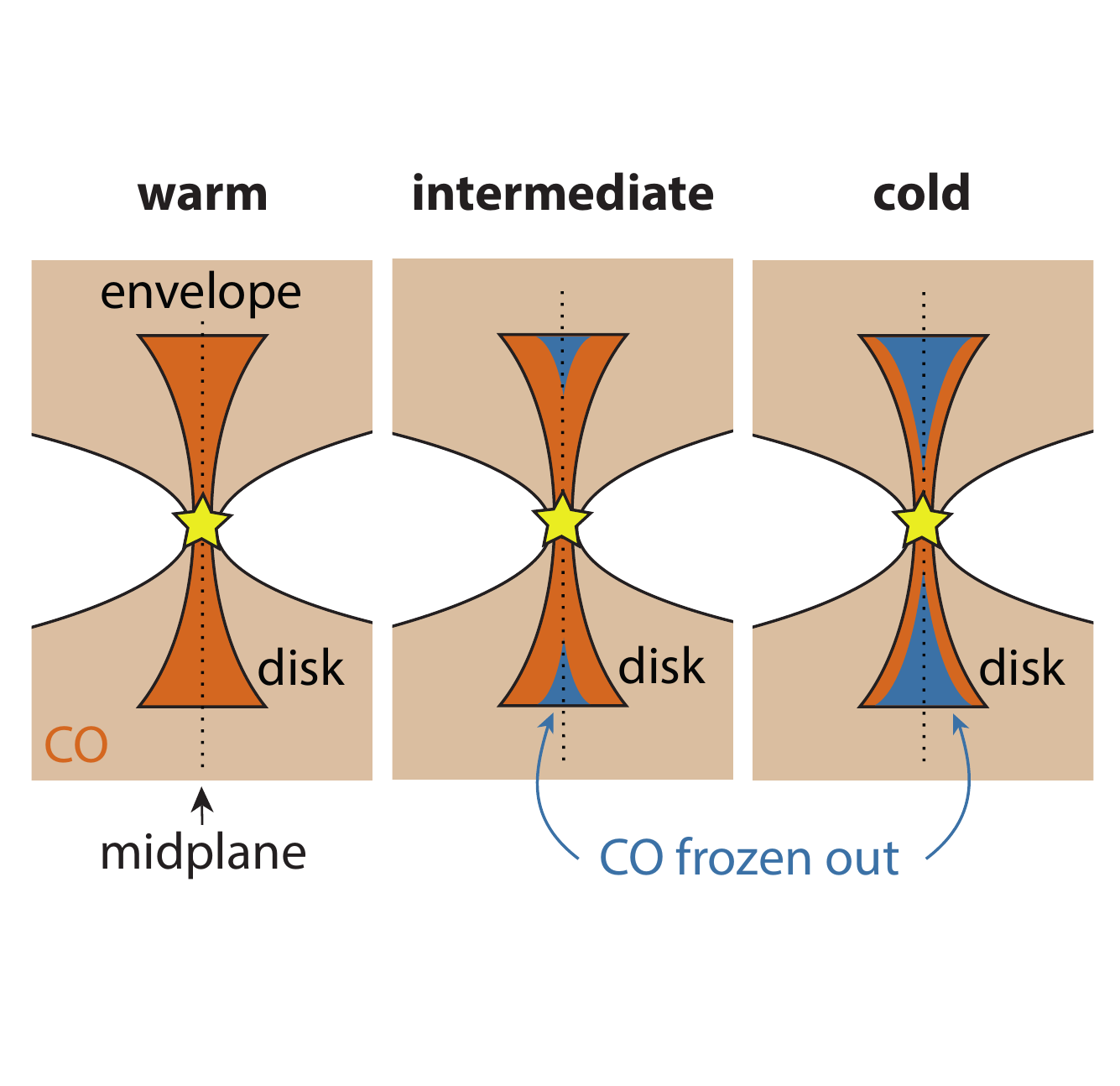}
\caption{Three different models for the CO distribution in embedded disks. \textit{Left panel:} a warm disk with no CO freeze out. \textit{Middle panel:} a slightly colder disk where CO is frozen out in the outer disk midplane. \textit{Right panel:} a cold disk where gaseous CO is only present in the inner disk and the disk surface layers. Gaseous CO is present in the inner envelope in all models. Figure reproduced from \citet{vantHoff2018b}.}
\label{fig:diskmodels}
\end{figure}

Figure~\ref{fig:diskmodels} shows a schematic overview of the warm, intermediate and cold disk models as presented by \citet{vantHoff2018b}. In the warm model, CO is present in the gas phase in the entire disk, whereas in the cold model CO is frozen out in most of the disk with gas-phase CO only present in the inner disk and disk surface layers. In the intermediate model CO freeze-out occurs in the outer midplane. A constant gas-phase CO abundance of 10$^{-4}$ with respect to H$_2$ is adopted in the regions where $T >$ 20 K. If the envelope is included in the radiative transfer, gas-phase CO is present in the $T >$ 20 K region at an abundance of 10$^{-4}$ as well. For the physical structure (dust density and temperature) we adopt the model for L1527 from \citet{Tobin2013}, who modeled the disk continuum emission by fitting both the visibilities and images of 870 $\mu$m and 3.4 mm observations, the multi-wavelength SED and L$^{\prime}$ scattered light images with 3D radiative transfer modeling.

\begin{figure}
\centering
\includegraphics[trim={0cm 22cm 14cm 0cm},clip]{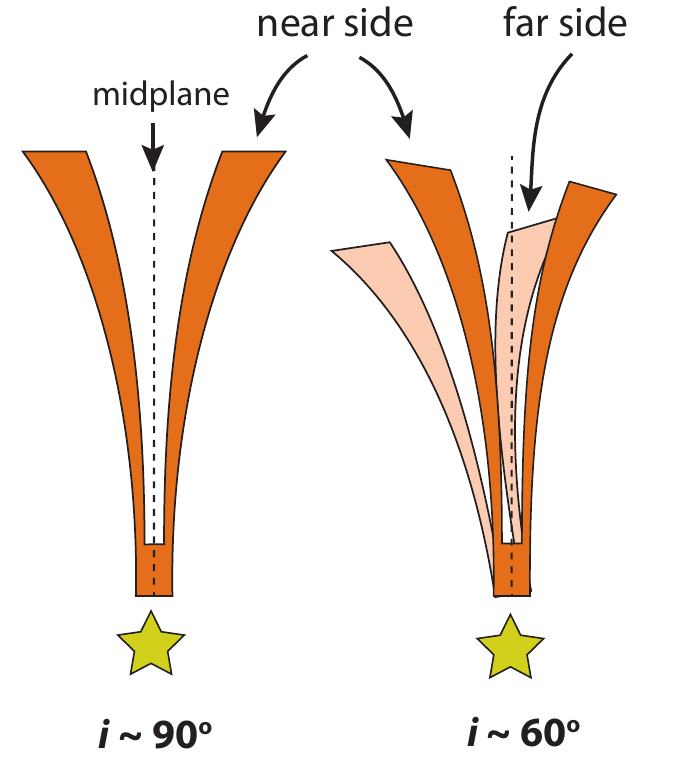}
\caption{Schematic representation of a disk with emission originating only in the surface layers viewed edge-on ($\sim90^\circ$; \textit{left panel} and at an inclination of $\sim60^\circ$ (\textit{right panel}). In the edge-on orientation, only the near side of the disk is visible and at sufficient angular resolution a V-shaped emission pattern is observed. In contrast, when the disk is $\sim60^\circ$ inclined, the far side of the disk becomes visible and emission from the far side appears to be coming from the midplane. This is especially problematic at low angular resolution, when the continuum disk is too small to map out the midplane or when the line is too weak to be detected in individual channels at high enough spectral resolution.}
\label{fig:incldisk}
\end{figure}

Figure~\ref{fig:incldisk} illustrates why observing freeze-out directly can be challenging in disks that are not viewed edge-on. 

\end{appendix}

\end{document}